\documentclass{emulateapj}
\usepackage{color}
\usepackage{natbib,graphicx,amsmath,amsthm,ulem,color}
\usepackage{latexsym, amssymb,longtable, epsf}
\newcommand{\be}{\begin{equation}}
\newcommand{\ee}{\end{equation}}
\newcommand{\beqn}{\begin{eqnarray}}
\newcommand{\eeqn}{\end{eqnarray}}
\newcommand{\bi}{\begin{itemize}}
\newcommand{\ei}{\end{itemize}}

\newcommand{\E}{Z_{\rm free}}
\def\refnew#1{(\ref{#1})}

\def\s{\, \rm s}
\def\K{\, \rm K}

\def\s{\, \rm s}
\def\km{\, \rm km}
\def\cm{\rm cm}

\def\g{\rm g}
\newcommand{\y}{}

\defcitealias{ttv1}{Paper I}	
\defcitealias{Batalhaetal12}{B12}	

\long\def\symbolfootnote[#1]#2{\begingroup\def\thefootnote{\fnsymbol{footnote}}
\footnote[#1]{#2}\endgroup}

\shorttitle{Density and Eccentricity of Kepler Planets}
\shortauthors{Wu \& Lithwick}

\begin{document}

\title{Density and Eccentricity of Kepler Planets}

\author{Yanqin Wu$^{1}$, Yoram Lithwick$^{2}$}
\affil{$^1$Department of Astronomy and Astrophysics, University of Toronto,  ON M5S 3H4, Canada;}
\affil{$^2$Department of Physics \& Astronomy, Northwestern University, Evanston, IL 60208, USA  \& Center for Interdisciplinary Exploration and Research in Astrophysics (CIERA)}


\begin{abstract}
  We analyze the transit timing variations (TTV) obtained by the
  {\it Kepler} mission for $22$ sub-jovian planet pairs 
  {\y (19 published, 3 new)} that lie close to mean motion resonances.
  We find that the TTV phases for most of these pairs lie close to
  zero, consistent with an eccentricity distribution that has a very
  low root-mean-squared value of $e \sim 0.01$; but about a quarter of
  the pairs possess much higher eccentricities, up to $e\sim 0.1 -
  0.4$.
  For the low-eccentricity pairs, we are able to statistically remove
  the effect of eccentricity to obtain planet masses from TTV data.
  These masses, together with those measured by radial velocity, yield
  a best fit mass-radius relation $M\sim 3 M_\oplus (R/R_\oplus)$.  This
  corresponds to a constant surface escape velocity of $\sim
  20\km/\s$.

  We separate the planets into two distinct groups, ``mid-sized''
  (those greater than $3 R_\oplus$), and ``compact'' (those smaller).
  All mid-sized planets are found to be less dense than water and
  therefore {\y must} contain extensive H/He envelopes
  {\y that are} comparable in mass to that of their cores. We argue
  that these planets have been significantly sculpted by
  photoevaporation. Surprisingly, mid-sized planets, a minority among
  Kepler candidates, are discovered exclusively around stars more
  massive than $0.8 M_\odot$.
  The compact planets, on the other hand, are often denser than
  water. Combining our density measurements with those from radial
  velocity studies, we find that hotter compact planets tend to be
  denser, with the hottest ones reaching rock density. Moreover,
  hotter planets tend to be smaller in size. These results can be
  explained if the compact planets are made of rocky cores overlaid
  with a small amount of hydrogen, $\leq 1\%$ in mass, with water
  contributing little to their masses or sizes.  Photoevaporation has
  exposed bare rocky cores in cases of the hottest planets.  Our
  conclusion that these planets are likely not water-worlds contrasts
  with some previous studies.

  While mid-sized planets most likely accreted their hydrogen envelope
  from the proto-planetary disks, compact planets could have obtained
  theirs via either accretion or outgassing.  The presence of the two
  distinct classes suggests that $3R_\oplus$ could be identified as
  the dividing line between `hot Neptunes' and `super-Earths.'
\end{abstract}



\section{Introduction}

The spectacular success of the {\it Kepler} mission \citep[][hereafter
B12]{Boruckiscience,Borucki,Batalhaetal12} opens our eyes to the world
of low-mass planets \citep[also see radial velocity discoveries,
e.g.][]{mayor-small}. The {\it Kepler} mission has uncovered a surprising
abundance of such planets close to their host stars and to each other.
While some planets are inferred to have high densities and therefore
are terrestrial-like, some have low densities indicating the presence
of a substantial gaseous envelope (see references in Table
\ref{tab:masses}). How do these low-mass planets form and migrate?
What gives rise to the range in planet radius and density?  What is
the internal composition of these planets? More measurements of planet
density may help resolving these puzzles. In particular, ascertaining
whether Kepler planets contain a significant amount of water would
shed light on the site of their construction \citep[see][for a
detailed discussion]{lopez}.

In addition to masses, the planets' eccentricities and inclinations
are valuable clues. Relative inclinations in Kepler systems are
inferred to be only a few degrees
\citep[e.g.,][]{tremaine,Fabryckyetal12,figura}. This flat
configuration invokes images of proto-planetary disks or planetesimal
disks.  Eccentricities, on the other hand, are relatively poorly
constrained \citep[see, e.g.][]{Moorhead}. Since eccentricities can be
excited by planet interactions and damped by planet-disk interactions
or tidal interactions with the host stars, they are a fossil record of
past dynamical events. In \citet{ttv1} (hereafter Paper1) we analyze
the TTV (transit time variation) signals of six planet pairs and show
that the TTV phases are consistent with  most planets
having in general small eccentricities. More
eccentricity determinations will help us rewind the clock and infer
the dynamical past of these planets.

With these two goals in mind, we analyze here the TTV signals for
pairs of transiting planetary candidates that lie near first-order
mean-motion resonances (MMR), using the Q0-Q6 {\it Kepler} table of transit
times in \citet{fordttv}. We also incorporate some results from
\citet{xie}. We then apply our analytical TTV expressions
\citepalias[derived in][]{ttv1} to infer planet properties. We
introduce a new element in this work by showing how true planet
density can be inferred from the nominal density measured by TTV, by
using statistical information from a large sample.  We outline our
sample selection criteria in \S \ref{sec:sample} and discuss our
results in \S \ref{sec:ecc},\ref{sec:density} \&
\ref{sec:correlation}. But first we recap some results from
\citetalias{ttv1} that are of relevance here.

For a pair of transiting planets that lie near a first order
mean-motion resonance, mutual gravitational perturbations cause each
planet's transit timing to vary sinusoidally with a sizable amplitude.
The periods, amplitudes, and phases of the sinusoids encode physical
properties of the planet pair:
\begin{itemize}
\item The periods of the two sinusoids are the same and we call
  them the {\it super-period}: 
\begin{equation}
 P^j = {1\over{ |{j/P'-(j-1)/P}|}} =  {{P'}\over{j|\Delta|}}\, ,
\label{eq:psuper}
\end{equation} 
where $P$ and $P'$ are orbital periods of the inner and outer planets,
respectively, $j$ stands for the $j:j-1$ resonance and $\Delta \equiv
{P'/P}(j-1)/j-1$ is the fractional distance to resonance. Most {\it Kepler}
pairs with clear TTV signals have $|\Delta|$ of order a few percent.

\item The amplitudes ($|V|$ and $|V|'$) are functions of both planet
  mass {\bf and} orbital eccentricity, roughly as (disregarding
  coefficients of order unity)
  \begin{eqnarray}
    {{|V|}\over{P}} & \sim &  {\mu'\over|\Delta|}\left(1+{|\E|\over|\Delta|}\right)\, , \nonumber \\
    {{|V'|}\over{P'}}& \sim &
    {\mu\over|\Delta|}\left(1+{|\E|\over|\Delta|}\right) \ .
\label{eq:ampp}
\end{eqnarray}
Here, $\mu$ is the mass ratio between the inner planet and the star
and $\mu'$ that of the outer planet.  $\E$ is a weighted sum of the
complex free eccentricity in both planets, 
\be
\E = f z_{\rm free} + g
z'_{\rm free}
\label{eq:zf}
\ee where $f$ and $g$ are sums of Laplace coefficients with
order-unity values, as listed in Table 3 of \citetalias{ttv1}.  The
free eccentricity, $z_{\rm free}$, is the excess in eccentricity over
what is forced by the resonance ('forced eccentricity').  We shall
often shorten `free eccentricity' to `eccentricity', since the forced
parts are mostly small for the pairs we consider ($\sim
10^{-3}$). These expressions show that amplitude ratio of the two
sinusoids is roughly related to the ratio of planet masses.  They also
highlight an inherent problem in TTV analysis: mass and eccentricity
are difficult to disentangle. However, if they could be
disentangled,\footnote{\y Radial velocity data, where available, can
  be used to break this degeneracy. Another method to do so, which we
  demonstrate in this paper, is to use the TTV pairs as an ensemble.}
TTV could be used to probe orbital eccentricity with a sensitivity of
a couple percent ($\sim |\Delta|$). This is superior to other
techniques like radial velocity or transit shape.

We define nominal masses for the two planets based on the observed TTV
amplitudes,
\begin{eqnarray}
m_{\rm nom}& \equiv & M_* \left|{V'\Delta\over
      P'g}\right|\pi j\, , \nonumber \\
m'_{\rm nom}& \equiv & M_* \left|{V\Delta\over Pf}\right|\pi
  j^{2/3}(j-1)^{1/3} \, . \label{eq:mvp} 
\end{eqnarray}
The true mass for the inner planet is $m=m_{\rm nom}/|1-Z_{\rm
  free}/(2g\Delta/3)|$, and analogously for the outer planet.  Hence
the nominal masses are typically upper limits: they are the true
masses at zero eccentricity, but usually lie above the true masses
when eccentricity is non-zero \citepalias{ttv1}.

\item Phases of the TTV sinusoids encode information about the free
  eccentricity. TTV phases are defined relative to the time when the
  longitude of conjunction points at the observer. The two phases
  ($\phi_{\rm ttv},\phi'_{\rm ttv}$) should lie at $(0,\pi)$ for zero
  eccentricity,\footnote{From now on we call the inner planet's
    $\phi_{\rm ttv}$ the TTV phase.}  while for most other
  eccentricities, the two phases can take other values but remain
  largely anti-correlated ($\phi'_{\rm ttv} \approx \phi_{\rm ttv} +
  \pi$). Equivalently, if the phases are close to $(0,\pi)$, then the
  pair likely have small eccentricity.  We show below that TTV phase
  can be used to infer planet eccentricity (\S \ref{sec:ecc}), and
  consequently, genuine planet masses (\S \ref{sec:density},).

\end{itemize}


\section{A sample of $22$ pairs}
\label{sec:sample}

Here we describe how our sample of $22$ TTV pairs are assembled.

Transit times for quarters 0-6 are published by \citet{fordttv} for a
large number of {\it Kepler} candidates.
We select from this list all pairs within $|\Delta|\leq 8\%$ from a
first-order MMR (2:1, 3:2, 4:3, 5:4).  For each pair, we obtain
average orbital periods and TTV amplitudes and phases using the
fitting procedure described in \citetalias{ttv1}.
We then calculate the periodogram for each TTV series and include a
pair for further consideration only when both planets have
periodograms that show a clear excess of power, by visual inspection,
at the desired super-period (eq. \ref{eq:psuper}).  We only include
pairs with super-period shorter than 1000 days, to ensure that more
than half of the TTV sinusoids have been observed. Amplitudes and
phases of the TTV sinusoids are measured at this super-period, using a
least-squares fit \citepalias{ttv1}.
Weeding out pairs whose sinusoid phases are not roughly out of phase
with each other, we find a total of $20$ pairs that pass these
thresholds, 6 of which are the confirmed systems analyzed in Paper I.

Compared to the exquisite TTV data for koi 137.01/02
\citep[Kepler-18c/d,][]{kepler18}, TTV data for typical {\it Kepler}
candidates are less accurate, since planets are typically smaller in
size and sampled at a lower cadence. Our stringent weeding criteria
are necessary to overcome random noise, and to isolate perturbations
from other planets. The latter is a prominent issue in {\it Kepler} data, as
planet pairs are often accompanied by other planets with near resonant
period ratios (e.g., KOI-82, KOI-500, KOI-2169).


In Fig. \ref{fig:ttvs}, we present the sinusoidal fits, error-bars,
and periodograms for $6$ of the above pairs, koi 156.01/03, koi
775.02/01, koi 841.01/02, koi 1215.01/02, koi 1241.02/01. 
{\y Koi 157.01/02, koi 775.02/01, koi 1241.02/01 have also been
  identified as TTV pairs by \citet{LissauerNature,Steffen}
  previously.}
For pairs that also appear in \citet{xie}, we adopt his TTV fits for
these systems as they are obtained using a longer baseline (through
quarter 9).
We also include $2$ new pairs from that work, koi 829.01/03, and koi
1336.01/02, that our shorter duration data failed to pick up. But we
do not use $3$ objects from that paper, {koi 877.01/02, koi 880.01/02
  and koi 869.03/02}: the first two because their TTV phases are not
roughly anti-correlated, which may indicate pollution, the last one
because of an unphysically large density (and mass) for the inner
planet: it requires a Jupiter mass for a $2.7 R_\oplus$ planet. It is
also likely contaminated by a nearby planet.

\begin{figure}
\centerline{\includegraphics[width=0.48\textwidth,trim=20 120 50 120,clip]{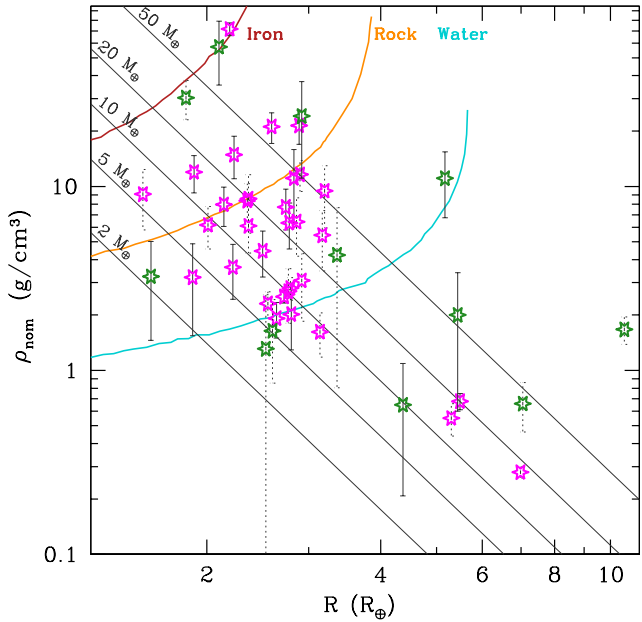}}
\caption{
  Nominal planet density versus radius for the TTV sample in Table
 \ref{tab:masses1}. Nominal densities are typically higher 
than true densities.  Magenta stars stand for previously
  confirmed pairs and green stars  those identified in this study
  (Fig. \ref{fig:ttvs}).   
 Error bars mark 68\% confidence limits on the nominal densities 
(solid error bars
  for inner planets and dashed for outer).
 Typical uncertainties in radius are 
$\sim 10\%$ and are not displayed here. The TTV method is currently sensitive 
to  planet masses as small as $2 M_\oplus$. The colored curves,
 marked with 'Iron', 'Rock' and 'Water' respectively, represent the 
theoretical mass-radius relation for these pure compositions  \citep{Fortney}.
}
\label{fig:ttv-density}
\end{figure}

So together with the $6$ planet pairs analyzed in \citetalias{ttv1},
we have a total of $22$ convincing TTV pairs 
{\y (19 published, 3 new)}.  TTV solutions for these pairs, including
their nominal 
{\y planet masses}, are listed in Table \ref{tab:masses1}, and
depicted in Fig. \ref{fig:ttv-density}.

\begin{table*}
\begin{center}
\begin{minipage}{180mm}
\caption{TTV analysis of $22$ planet pairs}
\begin{tabular}{|l|ccccc|ccc|ccc|}
  \hline
& 
  & \multicolumn{2}{c}{TTV amplitudes (min)} 
& \multicolumn{2}{c|}{TTV phases (deg)} 
& \multicolumn{3}{c|}{Inner Planet} & \multicolumn{3}{c|}{Outer Planet} \\
  KOI & $\Delta$ &
$|V|$ & $|V'|$  & $\phi_{\rm ttv}$  & $\phi'_{\rm ttv}$ 
& P &  $R$& $m_{\rm nom} (M_\oplus)$ 
& P' & $R'$ &$m'_{\rm nom} (M_\oplus) $ 
\\ \hline
%
  137.01/02 & $-.028$ & $  5.27(\pm  7\%)$ & $  4.06(\pm 10\%)$ & $  -4.3(\pm  4.0) $ & $ 168.7(\pm  5.0) $ & $  7.6420$ & $ 5.49$ & $  20.20\pm   2.02$ &$ 14.8590$ & $ 6.98$ & $  17.17\pm   1.20$ \\
  157.06/01 & $0.011$ & $ 12.99(\pm 40\%)$ & $  4.85(\pm 52\%)$ & $  97.3(\pm 25.2) $ & $ 237.0(\pm 34.3) $ & $ 10.3040$ & $ 1.89$ & $   3.87\pm   2.01$ &$ 13.0250$ & $ 2.92$ & $  13.69\pm   5.48$ \\
  157.03/04 & $-.027$ & $  7.10(\pm 48\%)$ & $ 20.55(\pm 68\%)$ & $  25.7(\pm 38.8) $ & $ 159.3(\pm 52.0) $ & $ 31.9960$ & $ 4.37$ & $   9.68\pm   6.58$ &$ 46.6890$ & $ 2.60$ & $   5.14\pm   2.47$ \\
  168.03/01 & $0.008$ & $ 43.98(\pm 37\%)$ & $ 18.95(\pm 23\%)$ & $ -70.1(\pm 19.3) $ & $ 123.6(\pm 18.2) $ & $  7.1070$ & $ 1.90$ & $  14.65\pm   3.37$ &$ 10.7420$ & $ 3.20$ & $  55.35\pm  20.48$ \\
  244.02/01 & $0.020$ & $  3.80(\pm 20\%)$ & $  1.04(\pm 36\%)$ & $   5.7(\pm 11.8) $ & $ 200.1(\pm 21.4) $ & $  6.2390$ & $ 2.80$ & $   7.91\pm   2.85$ &$ 12.7200$ & $ 5.30$ & $  14.59\pm   2.92$ \\
  870.01/02 & $0.013$ & $ 11.76(\pm 24\%)$ & $ 12.41(\pm 28\%)$ & $ -46.0(\pm 14.7) $ & $ 128.5(\pm 16.6) $ & $  5.9120$ & $ 2.50$ & $  12.44\pm   3.48$ &$  8.9860$ & $ 2.35$ & $  19.39\pm   4.65$ \\
  952.01/02 & $-.011$ & $  8.88(\pm 26\%)$ & $ 11.15(\pm 33\%)$ & $  45.2(\pm 15.5) $ & $ 227.8(\pm 18.3) $ & $  5.9010$ & $ 2.22$ & $   7.11\pm   2.34$ &$  8.7520$ & $ 2.01$ & $   8.94\pm   2.32$ \\
 1102.02/01 & $0.009$ & $ 41.03(\pm 19\%)$ & $ 37.60(\pm 25\%)$ & $  -3.9(\pm  9.5) $ & $ 189.5(\pm 14.3) $ & $  8.1460$ & $ 2.74$ & $  28.39\pm   7.10$ &$ 12.3330$ & $ 2.90$ & $  50.58\pm   9.61$ \\
\hline
  148.01/02 & $0.012$ & $  4.00(\pm 27\%)$ & $  3.31(\pm 24\%)$ & $ -23.5(\pm 17.6) $ & $ 199.5(\pm 18.5) $ & $  4.7780$ & $ 2.14$ & $  13.96\pm   3.35$ &$  9.6740$ & $ 3.14$ & $   8.93\pm   2.41$ \\
$^*$  152.03/02 & $0.016$ & $  7.92(\pm 33\%)$ & $ 26.21(\pm 19\%)$ & $ -29.6(\pm 28.6) $ & $ 123.2(\pm  8.7) $ & $ 13.4850$ & $ 2.59$ & $  65.59\pm  12.46$ &$ 27.4030$ & $ 2.77$ & $  10.22\pm   3.37$ \\
  248.01/02 & $0.010$ & $  9.22(\pm 16\%)$ & $ 20.02(\pm  8\%)$ & $  54.1(\pm  9.3) $ & $ 218.0(\pm  5.5) $ & $  7.2040$ & $ 2.72$ & $   9.07\pm   0.73$ &$ 10.9130$ & $ 2.55$ & $   6.83\pm   1.09$ \\
  500.01/02 & $0.012$ & $  8.50(\pm 14\%)$ & $  7.06(\pm 22\%)$ & $  -9.1(\pm  9.0) $ & $ 178.0(\pm 13.3) $ & $  7.0530$ & $ 2.64$ & $   6.31\pm   1.39$ &$  9.5220$ & $ 2.79$ & $  10.88\pm   1.52$ \\
$^*$  829.01/03 & $0.034$ & $ 16.70(\pm 36\%)$ & $ 22.32(\pm 21\%)$ & $ -33.9(\pm 11.1) $ & $ 135.9(\pm 13.0) $ & $ 18.6490$ & $ 2.89$ & $  92.43\pm  19.41$ &$ 38.5590$ & $ 3.17$ & $  30.90\pm  11.12$ \\
  898.01/03 & $0.028$ & $  7.34(\pm 29\%)$ & $ 10.80(\pm 43\%)$ & $  63.8(\pm 15.0) $ & $ 213.2(\pm 23.3) $ & $  9.7710$ & $ 2.83$ & $  44.89\pm  19.30$ &$ 20.0900$ & $ 2.36$ & $  14.34\pm   4.16$ \\
$^*$ 1270.01/02 & $0.013$ & $  3.17(\pm 36\%)$ & $ 37.15(\pm  8\%)$ & $  96.6(\pm 18.5) $ & $ 281.7(\pm  4.7) $ & $  5.7290$ & $ 2.19$ & $ 134.49\pm  10.76$ &$ 11.6090$ & $ 1.55$ & $   6.03\pm   2.17$ \\
 1336.01/02 & $0.016$ & $ 18.00(\pm 35\%)$ & $ 26.93(\pm 28\%)$ & $ -61.7(\pm 17.3) $ & $ 125.6(\pm 15.8) $ & $ 10.2190$ & $ 2.78$ & $  24.33\pm   6.81$ &$ 15.5740$ & $ 2.86$ & $  26.86\pm   9.40$ \\
 1589.01/02 & $-.016$ & $ 12.82(\pm 35\%)$ & $ 29.38(\pm 26\%)$ & $  21.3(\pm 26.3) $ & $ 184.6(\pm 14.1) $ & $  8.7260$ & $ 2.23$ & $  29.45\pm   7.66$ &$ 12.8820$ & $ 2.36$ & $  20.12\pm   7.04$ \\
\hline
  156.01/03 & $-.024$ & $  2.07(\pm 97\%)$ & $  2.00(\pm 55\%)$ & $   3.7(\pm 74.8) $ & $ 168.4(\pm 33.9) $ & $  8.0410$ & $ 1.60$ & $   2.37\pm   1.30$ &$ 11.7760$ & $ 2.53$ & $   3.79\pm   3.68$ \\
$^*$  775.02/01 & $0.040$ & $ 11.81(\pm 24\%)$ & $ 13.94(\pm 38\%)$ & $-153.1(\pm 12.7) $ & $  40.6(\pm 23.2) $ & $  7.8770$ & $ 2.10$ & $  94.79\pm  36.02$ &$ 16.3850$ & $ 1.84$ & $  33.78\pm   8.11$ \\
  841.01/02 & $0.021$ & $ 29.38(\pm 30\%)$ & $ 20.13(\pm 70\%)$ & $  30.5(\pm 13.2) $ & $ 198.1(\pm 28.6) $ & $ 15.3360$ & $ 5.44$ & $  57.49\pm  40.24$ &$ 31.3280$ & $ 7.05$ & $  41.59\pm  12.48$ \\
$^*$ 1215.01/02 & $-.047$ & $  9.86(\pm 81\%)$ & $ 25.93(\pm 54\%)$ & $  29.9(\pm 56.7) $ & $ 183.8(\pm 29.1) $ & $ 17.3240$ & $ 2.92$ & $ 107.18\pm  57.88$ &$ 33.0060$ & $ 3.36$ & $  28.74\pm  23.28$ \\
$^*$ 1241.02/01 & $0.019$ & $149.21(\pm 17\%)$ & $ 58.46(\pm 39\%)$ & $ 125.9(\pm 12.3) $ & $ 360.7(\pm 13.1) $ & $ 10.5040$ & $ 5.17$ & $ 273.59\pm 106.70$ &$ 21.4010$ & $10.57$ & $ 353.11\pm  60.03$ \\
\hline
\end{tabular}
\tablecomments{A list of our sample of $22$ TTV pairs.  
  {\y Here, $\Delta$ is the distance to the nearest first order MMR,
    $|V|$ is the measured TTV amplitude (in }
  {\y minutes) for the inner planet, $\phi_{\rm ttv}$ its TTV phase
    ($1-\sigma$ error bars in parantheses), $P$ its orbital period (in
    days), $R$ its radius (in $R_\oplus$) and $m_{\rm nom}$ its
    nominal mass (in $M_\oplus$). The primed quantities are the same
    but for the outer planets.}
  The first $8$ pairs are confirmed systems (corresponding to Kepler
  18c/d, 11b/c, 11e/f, 23 b/c, 25 b/c, 28 b/c, 32 b/c, 24 b/c,
  respectively), six of which were analyzed in \citetalias{ttv1}.
  TTV parameters for the next $8$ pairs are adopted from \citet{xie},
  and TTV sinusoids for the last $5$ pairs are reported 
{\y in} Fig. \ref{fig:ttvs}. 
{\y See \S \ref{subsec:starsize} for the stellar parameters we adopt.}
Pairs marked with the $^*$ sign are suspected to have significant free
eccentricity (\S \ref{sec:ecc}). Typical error bars in planetary radii
are of order $10\%$, smaller than those in TTV amplitudes. These are
not included in calculating uncertainties in nominal densities.  }
%
\label{tab:masses1}
\end{minipage}
\end{center}
\end{table*}

\subsection{Selection Effects}
\label{subsec:selection}

There are $151$ planet pairs in the \citetalias{Batalhaetal12} catalog
that lie within $|\Delta| \leq 5\%$ from a first-order MMR.  Our
sample of $22$ pairs is $\sim 15\%$ of the total.  Since the strengths
of TTV signals depend on planet mass and eccentricity, it is of
concern whether our sample is biased towards planets that have higher
density or higher eccentricity 
{\y when compared against planets {\bf{of the same sizes and at the
      same periods}}}.
Moreover, we sample only planets near MMRs, which may in principle
have different properties than those far from resonance.

We argue here that, in comparison with the B12 catalog, our sample
favours larger planets,
as well as planets that have {\y a particular range of} orbital
periods, but not necessarily planets with higher density or
eccentricity. First, larger planets show deeper transits and allow
more precise timing.
The median size of the \citetalias{Batalhaetal12} pairs is $2.18 R_\oplus$
while our sample has a significantly larger median of $2.72 R_\oplus$.
%
Second, our TTV sample excludes planets with long orbital periods and
hence long super-periods. It also excludes planets too close to their
stars, which have smaller intrinsic TTV amplitudes and less accurate
TTV measurements due to their shorter transit duration.  The median
inner planet period in our sample is $8.04$ days, with a standard
deviation from the median of $5.98$ days.
In contrast, values for the \citetalias{Batalhaetal12} sample are $7.88$
days and $14.1$ days. This means that we are mostly probing pairs
clustered around $\sim 8$ day orbital periods.

These two biases alone could explain why our sample is only $15\%$ of
the overall sample of candidate pairs. 
In fact, when restricting to pairs similar in properties to our sample
(inner periods between $5$ and $20$ days, planet radii greater than
$1.7 R_\oplus$, $|\Delta|<5\%$), we find $48$ pairs in the
\citetalias{Batalhaetal12} catalog.  This leaves little room to
suspect that we are significantly biased toward higher density and
eccentricity.

A related issue concerns sensitivity in nominal mass. Currently Kepler
can probe TTV signals with amplitude upward of $\sim 5$
minutes. Adopting $|\Delta| \approx 2\%$, and an inner period of $8$
days, this implies that TTV's are sensitive to planets with nominal
masses upward of $5 M_\oplus$, around a sun-like star. Sensitivity
around lower-mass stars is even better, $\sim 2 M_\oplus$
(Fig. \ref{fig:ttv-density}), thanks to their deeper transit at the
same planetary size. Lastly, we find that, with a factor of $2$ longer
observation, or moderate improvement in transit timing precisions, one
should be able to detect TTV sinusoids in many more pairs than
reported here.

We argue that our near-resonant pairs are not special. First, planet
pairs in the \citetalias{Batalhaetal12} sample show similar size
distribution, whether they are close to an MMR or not.
Moreover, while our TTV sample is restricted to planets near
first-order MMRs, the RV planets we include in our analysis below are
not, and those appear to have similar properties.

\subsection{Uncertainties in Stellar Parameters}
\label{subsec:starsize}

{\y Prompted by the referee, we discuss uncertainties in the stellar
  parameters (the most important being stellar radii) that we adopt.
}
  
{\y There are two groups of Kepler host stars for which stellar radii
  can contain large uncertainties.  The first group are stars that are
  hotter than $5400{\rm \,K}$ -- their stellar $\log g$ values from
  KIC \citep{KIC} are fairly uncertain as these stars can be subgiants
  impersonating main-sequence stars. In this case, their radii can be
  underestimated by factor of $1.5-2$, leading to a corresponding
  underestimate in planet radii and an overestimate, by a factor of
  $3-8$, in planet density.  }

 {\y The second group contains M-dwarfs (stellar mass below $0.45
  M_\odot$) whose radii are notoriously difficult to determine, due to
  a lack of good evolutironary tracks \citep[see,
  e.g.][]{Muirhead,Dressing}. Fortunately, none of our stars are
  M-dwarfs.}


{\y Of the $21$ host stars that we study, $11$ have had reliable
  stellar radii measurements. These include $8$ systems that have
  updated $\log g$ measurements from spectroscopy (B12): koi
  137,148,156,157,1215,500,841,and 898}{\y , all of which} {\y are
  main-sequence stars and not subgiants (private communication,
  Huber);} {\y and $3$ systems that have asteroseismologically
  determined masses and radii \citep[koi 168,244,1241][]{Huber}.  }

{\y For the remaining $10$ stars, we}{\y adopt B12 values for the
  stellar radii, which are inherited from the KIC values and}{\y may
  contain errors. Six out of this group (koi 248, 952, 870, 1589,
  1102, 1336)
  are considered to harbour low-eccentricity planets (see below) and
  they contribute to our final conclusion. Among these, new radii
  determinations (that concur with KIC values) for koi 248,952 are
  provided by \citet{Dressing}. Koi 870, 1102 and 1336 are confirmed
  systems, named as Kepler-28, Kepler-24 and Kepler-58, respectively. 

  So while more secure stellar parameters are desirable for our
  systems, we believe that our basic conclusions}
{\y are unlikely to be affected by their uncertainties.  }



\section{Measuring Planet Eccentricities}
\label{sec:ecc}

\begin{figure}
\centerline{\includegraphics[width=0.48\textwidth,trim=20 150 60 120,clip]{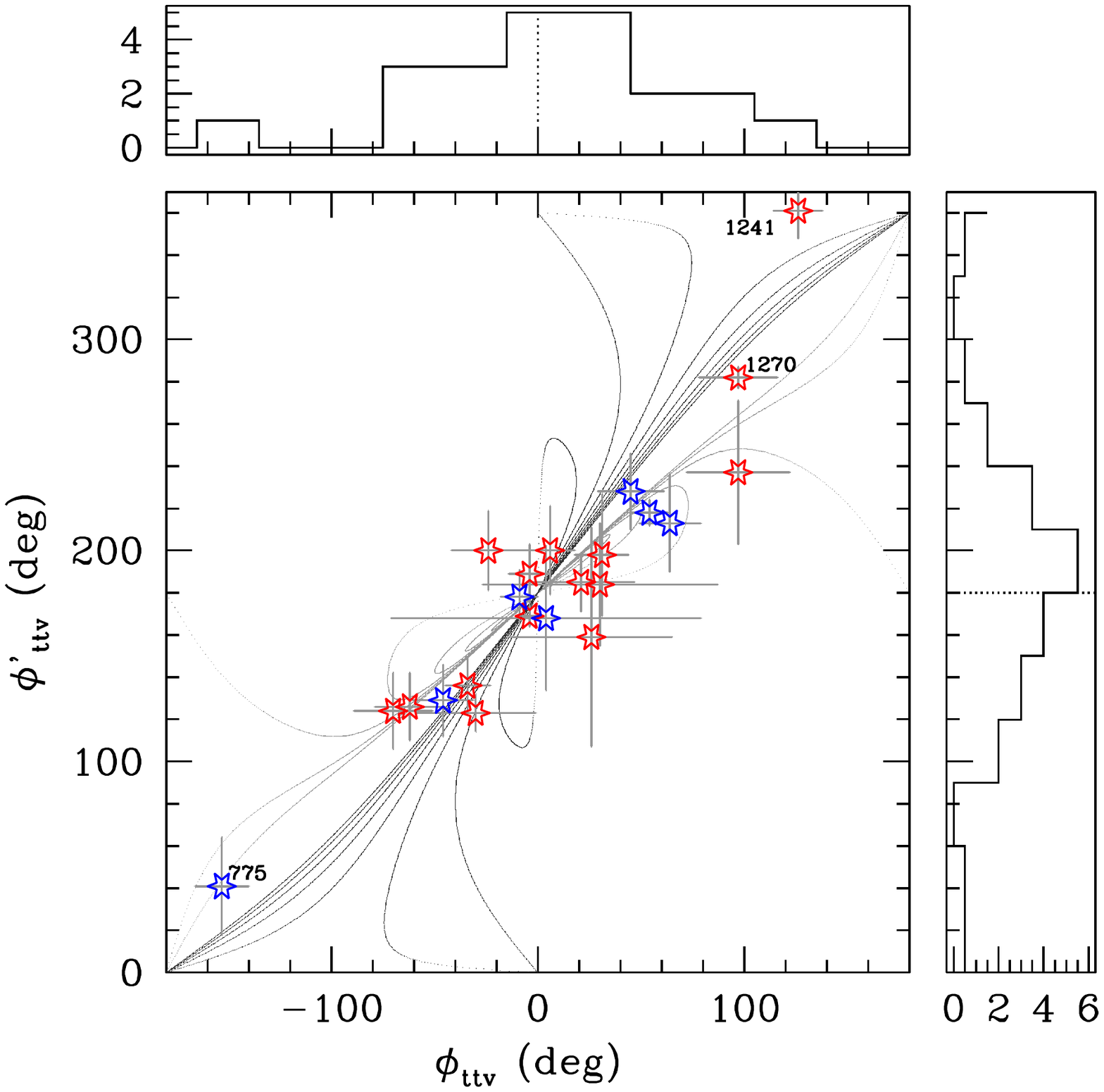}}
\caption{
  Phases of the TTV sinusoids for the inner ($\phi_{\rm ttv}$) and
  outer ($\phi'_{\rm ttv}$) planets for our $22$ pairs.  Pairs with
  very small eccentricity ($\ll |\Delta|$) would lie near
  $(0,180^o)$. For higher eccentricity, phases should lie along the
various grey curves (from $0.25 |\Delta|$ to $2 |\Delta|$ in
increments of $0.25 |\Delta|$, dark grey for 2:1 MMR and lighter grey
for 3:2), becoming increasingly evenly distributed along the diagonal
line, which corresponds to anti-correlated phases.
 Histograms for the measured $\phi_{\rm ttv}$ and
 $\phi'_{\rm ttv}$ are displayed at the top and 
 side panels, respectively.  The measured TTV phases  
cluster around $(0,180^o)$. 
 In this and all subsequent figures,  
red  denotes planets
 around stars more massive than $0.8 M_\odot$, 
and blue  those around less massive stars.
  Error-bars indicate 
 68\% confidence limit.
Phases of three pairs (labelled with their koi numbers) 
deviate significantly
 from zero.  These also  exhibit
anomalously high nominal
  densities (Table \ref{tab:masses1}), 
suggesting eccentricity $\gtrsim |\Delta|$.
The pair koi 157.06/01 (Kepler 11b/c) also has high phases (the red
point at $(97^o,237^o)$), but this likely results from 
 its dominant resonance (5:4) being close to other nearby resonances 
(4:3, 6:5, etc).
}
\label{fig:ttv-phase}
\end{figure}

\begin{figure}
\centerline{\includegraphics[width=0.49\textwidth,trim=20 100 40 115,clip]{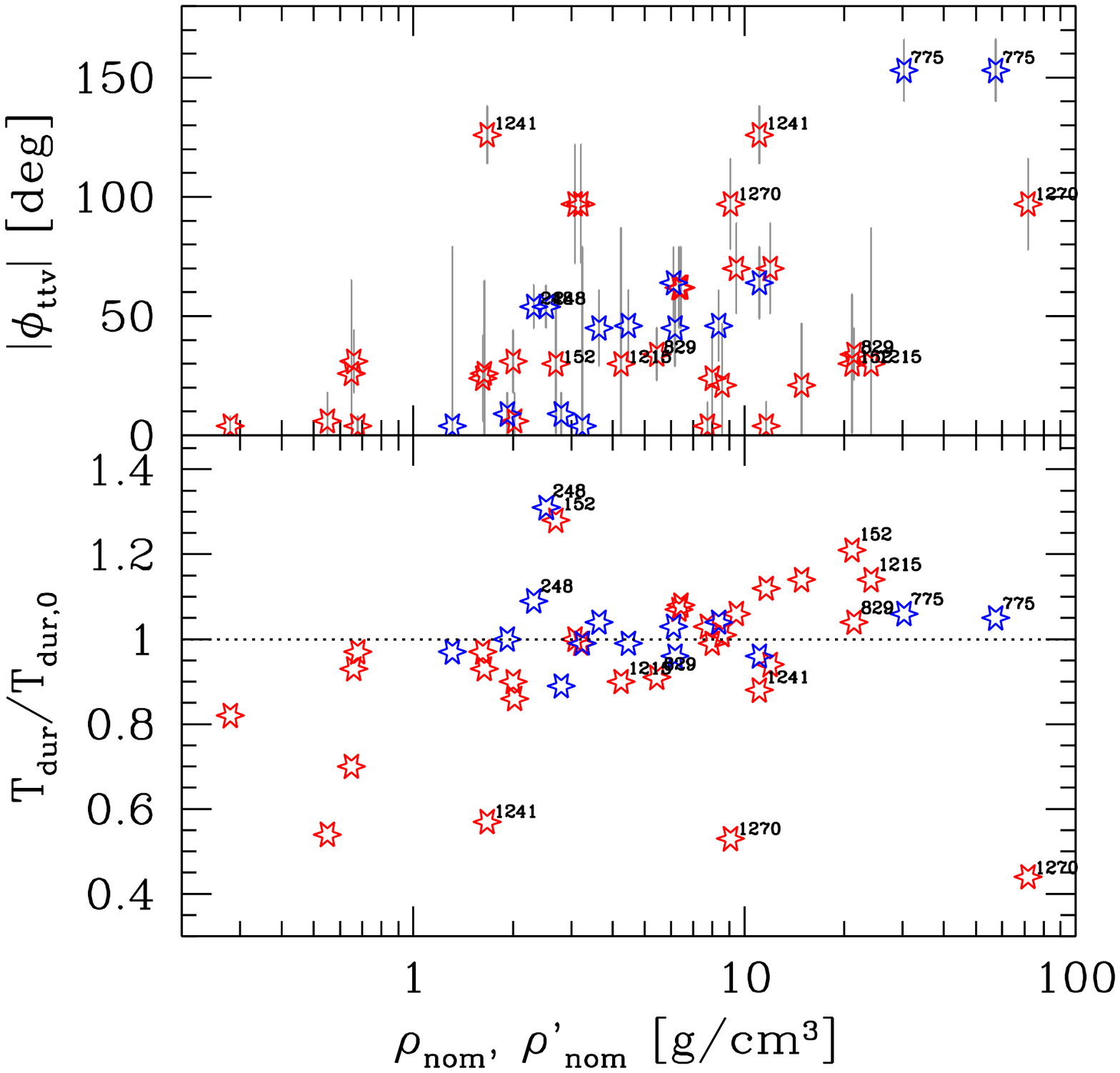}}
\caption{
{\y Identifying} pairs 
with high eccentricity by their transit
  duration, nominal densities, and TTV phases. 
 We plot TTV
  phase ($\phi_{\rm ttv}$, top panel) and transit duration (lower
  panel) versus nominal density for all $22$ pairs.  The transit
  duration is normalized by $T_{\rm dur,0}$,   the duration of a
  circular orbit with zero impact parameter. All circular orbits should lie
  below the line of unity, while eccentric orbits can lie above or below.
  Three pairs (Koi 775, 1241, 1270, marked out in
  Fig. \ref{fig:ttv-phase}) exhibit high nominal densities, large TTV
  phases, and/or longer than unity transit duration.  These systems
  almost certainly have high free eccentricities ($\gtrsim$ a few
$\times \Delta$).
Three pairs (Koi 152, 829, 1215) have low TTV phases. But their high
nominal densities, and in many cases, longer than unity transit
duration, suggest that they possess high eccentricity. Together, these
$6$ pairs constitute our high-e population.
Transits of koi 248 last longer than expected of circular orbits.  We
suspect radius determination for the host star is too small by
$10-30\%$ (but we neglect this in the following analysis).}
\label{fig:rhophi}
\end{figure}

One of the most important results from \citetalias{ttv1} is the
utility of the TTV phase ($\phi_{\rm ttv}$) for inferring the value of
the free eccentricity.  If a significant fraction of systems have
$\phi_{\rm ttv} \approx 0$, it can be inferred that most of those have
$|\E|\lesssim |\Delta|$.  In \citetalias{ttv1}, we find that  the
  TTV phases of the six analyzed pairs are clustered near zero,
  indicating that in general the free eccentricity is small ($|\E|
  \leq \Delta$).

  Now armed with $22$ pairs, we explore the eccentricity distribution
  further. We find that for this larger sample too the
  TTV phases are not uniformly distributed, but cluster around zero
  (Fig. \ref{fig:ttv-phase}), again indicating that the general
  population possesses small free eccentricity. However, there are a
  number of pairs that buck this trend. These are marked in
  Fig. \ref{fig:rhophi}, based on three indices: TTV phase, nominal
  density, and transit duration. Both the TTV phases and nominal
  densities (compared to other planets of similar sizes) for Koi 775,
  1241, 1270 are large, suggesting high eccentricities.
  Since for the high-e population, $\phi_{\rm ttv}$ is uniformly
  distributed between $(-\pi, \pi)$, we expect to find pairs that have
  high-e but low TTV phases. And indeed, pairs Koi 152, 829, 1215,
  while showing small phases, have high nominal densities. The transit
  durations for these planets are often longer than expected for
  circular orbits.  We categorize these $6$ pairs as the high-e
  population and analyze them in \S \ref{subsec:eccdiscuss}.  It is
  interesting to note that all planets with nominal densities above
  $\sim 15\g/\cm^3$ are now classified as high-e. {\y The true
    densities of these planets are lower than the nominal values by
    some factors, depending on the actual eccentricities
    (eq. \ref{eq:ampp}). Unfortunately, our current strategy can not
    uncover the actual eccentricities for these planets, only lower
    limits.}

\begin{figure}
\centerline{\includegraphics[width=0.48\textwidth,trim=25 120 20 100,clip]{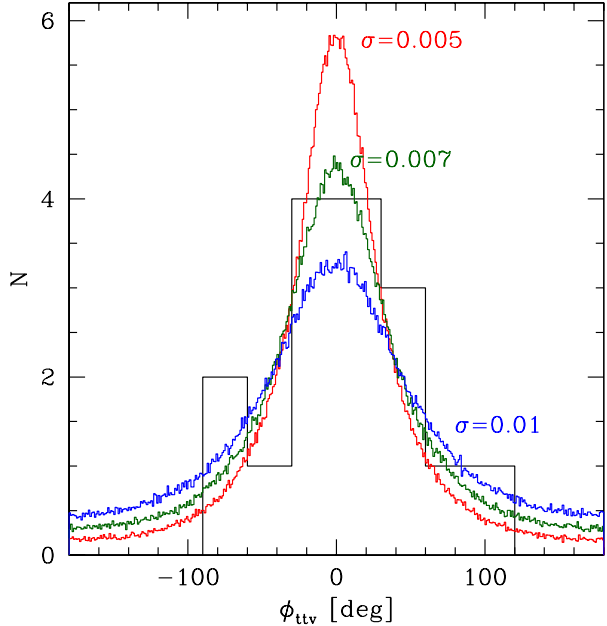}}
\caption{The distribution of TTV phases. The black histogram is the
  observed distribution for the $16$ low-e pairs. The colored curves
  are simulation results at different values of eccentricity
  dispersion (Eq. \ref{eq:psigma}), calculated for a mock catalog that
  has the same period ratios as the observed one.  The curve with
  $\sigma=0.007$ best reproduces the observation, with an uncertainty
  of $\sim 20\%$. The RMS eccentricity of individual planets is
  $\sqrt{2} \sigma \approx 0.01$.}
\label{fig:sigma}
\end{figure}

  The remaining $16$ pairs ($\sim 75\%$) are consistent with low
  eccentricity.  We assume them to be a homogeneous population
  satisfying a single eccentricity distribution. TTV phases depend on
  $\E$, a weighted sum of the two planets' free eccentricities
  (Eq. \ref{eq:zf}).  Splitting into real and imaginary parts,
\begin{equation}
z_{\rm free} = e_x + i e_y\, ,\label{eq:xy}
\end{equation} 
and adopting a 1-D gaussian distribution for these components (for
both the inner and outer planets) , $P(e_x) = P_\sigma(e_x)$, $ P(e_y)
= P_\sigma(e_y)$, with
\begin{equation}
P_\sigma(e_x) = {1\over{\sqrt{2 \pi}}\sigma} \exp(- {{e_x^2}\over{2\sigma^2}})\, ,
\label{eq:psigma}
\end{equation}
we can calculate the resulting distribution of TTV phases.  We assume
$e_x$ and $e_y$ are uncorrelated, 
i.e., that the phases of the complex eccentricities are random, as
they would be due to secular or GR precession \citepalias{ttv1}. TTV
phases are either uniformly distributed or clustered around zero,
depending on the relative ratio of $\sigma/|\Delta|$, and on the
resonance of relevance. Generating a mock catalogue that has the same
period ratios as our low-e sample, we produce different phase
distributions for different values of $\sigma$
(Fig. \ref{fig:sigma}). The current sample lead us to conclude that a
value of $\sigma \approx 0.007$ provides the best fit, with an
uncertainty in $\sigma$ of order $20\%$.  This value corresponds to a
RMS value in eccentricity ($e = \sqrt{e_x^2 + e_y^2}\,$) of $0.01$. So
it appears that the low-e population has free eccentricities that are
of order (but somewhat smaller than) the typical resonance offset
($|\Delta| \sim 0.02$). \footnote{We have also experimented with other
  forms of eccentricity distribution, e.g., a $\delta$-function plus a
  Gaussian, 1-D Gaussian, etc., but find no improved fit to the data.}

This exercise demonstrates the exceptional sensitivity of TTV
measurements to even small values of eccentricity. We discuss the
implication of these results in \S \ref{subsec:eccdiscuss}.

\section{ True Densities}
\label{sec:density}

Employing our newly found eccentricity distribution, we perform
statistical simulations to correct for the eccentricity effect, and
thereby obtain true planet densities for the low-e population, subject
to statistical and measurement errors. Densities of the high-e sample
are not recoverable.


\begin{table}
\begin{center}
\begin{minipage}{80mm}
\caption{Density after Correcting for Eccentricity Effect}
\begin{tabular}{|lcc|lcc|c|}
  \hline
KOI & $\rho (\g/\cm^3)$ & $\rho' (\g/\cm^3)$ & ${{\rho}\over{\rho_{\rm nom}}}$ & ${{\rho'}\over{\rho'_{\rm nom}}}$ &
${{e+e'}\over{2}}$ \\
 \hline
  137.01/02 & $ 0.4^{+ 74\%}_{- 52\%}$ & $ 0.2^{+ 29\%}_{- 28\%}$ & $0.66$ &  $0.82$ & $0.007$ \\
  157.06/01 & $ 3.2^{+ 52\%}_{- 52\%}$ & $ 3.1^{+ 40\%}_{- 40\%}$ & $1.00$ &  $1.00$ & $0.000$ \\
  168.03/01 & $ 5.6^{+ 96\%}_{- 65\%}$ & $ 3.7^{+107\%}_{- 73\%}$ & $0.47$ &  $0.40$ & $0.008$ \\
  244.02/01 & $ 0.9^{+ 96\%}_{- 70\%}$ & $ 0.4^{+ 38\%}_{- 37\%}$ & $0.43$ &  $0.72$ & $0.007$ \\
  870.01/02 & $ 2.6^{+ 67\%}_{- 54\%}$ & $ 4.2^{+ 73\%}_{- 55\%}$ & $0.57$ &  $0.50$ & $0.008$ \\
  952.01/02 & $ 1.9^{+ 76\%}_{- 59\%}$ & $ 2.8^{+ 82\%}_{- 59\%}$ & $0.51$ &  $0.45$ & $0.008$ \\
 1102.02/01 & $ 3.3^{+ 80\%}_{- 57\%}$ & $ 4.4^{+ 90\%}_{- 59\%}$ & $0.43$ &  $0.38$ & $0.008$ \\
  148.01/02 & $ 2.2^{+ 95\%}_{- 63\%}$ & $ 0.9^{+ 53\%}_{- 47\%}$ & $0.27$ &  $0.55$ & $0.008$ \\
  156.01/03 & $ 2.6^{+ 82\%}_{- 68\%}$ & $ 1.0^{+123\%}_{-107\%}$ & $0.80$ &  $0.76$ & $0.007$ \\
  157.03/04 & $ 0.5^{+ 87\%}_{- 78\%}$ & $ 1.3^{+ 82\%}_{- 64\%}$ & $0.83$ &  $0.79$ & $0.008$ \\
  248.01/02 & $ 1.3^{+ 90\%}_{- 59\%}$ & $ 1.1^{+105\%}_{- 65\%}$ & $0.53$ &  $0.47$ & $0.008$ \\
 1336.01/02 & $ 4.7^{+ 77\%}_{- 58\%}$ & $ 4.2^{+ 89\%}_{- 65\%}$ & $0.74$ &  $0.65$ & $0.008$ \\
  500.01/02 & $ 1.0^{+ 76\%}_{- 53\%}$ & $ 1.3^{+ 85\%}_{- 54\%}$ & $0.52$ &  $0.48$ & $0.007$ \\
  841.01/02 & $ 0.6^{+ 81\%}_{- 79\%}$ & $ 0.4^{+ 36\%}_{- 37\%}$ & $0.32$ &  $0.63$ & $0.009$ \\
  898.01/03 & $ 3.7^{+ 49\%}_{- 50\%}$ & $ 4.7^{+ 32\%}_{- 34\%}$ & $0.33$ &  $0.77$ & $0.012$ \\
 1589.01/02 & $ 9.6^{+ 75\%}_{- 53\%}$ & $ 5.1^{+ 93\%}_{- 63\%}$ & $0.65$ &  $0.60$ & $0.007$ \\
\hline
\end{tabular}
\tablecomments{Planet densities after applying the statistical model
  for the eccentricity distribution. The column $\rho$ lists the
  median density for the inner planet (and primed quantities for the
  outer planet), a product of the median correction factor
  ($\rho/\rho_{\rm nom}$) and the nominal density. Error-bars on the
  correction factor reflect the width within which $68\%$ of all
  possible solutions fall. Error-bars on the median density are a
  quadratic sum of uncertainties in the correction factor and
  uncertainties in nominal mass (measurement error in TTV amplitude).
  The last column lists the average of the two median
  eccentricities, as a rough indicator for the magnitude of
  eccentricity in the pair.
  Our values for Koi 137.01/02 (Kepler 18c/d) agree with previous
  determinations of $\rho=0.59\pm 0.07$, $\rho'=0.27\pm 0.03$
  \citep{kepler18}. We do not perform a correction for Koi 157.06/01
  (Kepler 11b/c) as its TTV phase is polluted by other resonances.
%
}
 \label{tab:masses2}
\end{minipage}
\end{center}
\end{table}

\begin{table}[b]
\begin{center}
\begin{minipage}{85mm}
\caption{Literature Determinations of Planet Density}
%
\begin{tabular}{|clll|l|}
  \hline
  Planet & $R $ &$M $&  $\rho $  & Reference  \\
   & $(R_\oplus)$ & $(M_\oplus)$ &   $(\g/\cm^3)$ & \\
 \hline
GJ1214b &   $2.68$&$6.55$ & $1.87 \pm 0.4$ & \citet{gj1214b}\\
GJ3470b &   $4.1$ &$14.5$ & $1.16 \pm 0.17$ & \citet{gj3470} \\
GJ436b &    $4.0$ &$24.3$ & $2.09 \pm 0.14$ & \citet{gj436} \\
Kepler-4b&  $3.99$&$24.5$ & $1.9 \pm 0.4$ & \citet{kepler4b}\\
Kepler-10b& $1.42$&$4.56$ & $8.8\pm 2.5$ & \citet{kepler10b}\\
HAT-P-26b & $5.8$ &$18$   & $0.4 \pm 0.1$ & \citet{hatp26}\\
HAT-P-11b & $4.9$ &$26$   & $1.2 \pm 0.13$ & \citet{hatp11}\\
CoRoT-7b &  $1.6$ &$7.4$  & $10.4\pm 1.8$ & \citet{corot7b}\\
CoRoT-8b & $6.4$ & $68.7$ & $1.6\pm 0.1$ & \citet{Borde10} \\
55 Cnc e &  $2.2$ &$8.6$  & $5.9\pm 1.3$ & \citet{endl}\\
& & & & \citet{55cnc}\\
kepler-20b& $1.9$ &$8.7$  & $7.1 \pm 1.8$ & \citet{kepler20}\\
kepler-20c& $3.1$ &$16.1$ & $3.0 \pm 0.8$ & \citet{kepler20}\\
Kepler-36b& $1.5$ &$4.5$  & $7.5\pm 0.7$ & \citet{kepler36}\\
Kepler-36c& $3.7$ &$8.1$  & $0.89\pm 0.06$ & \citet{kepler36}\\
  \hline
\end{tabular}
\tablecomments{Transiting planets with mass determinations using the
  radial velocity technique (except for Kepler-36b/c which is through
  TTV). Here, we only include planets that are smaller than $9
  R_\oplus$ in size.
}
\label{tab:masses}
\end{minipage}
\end{center}
\end{table}

For each planet pair, we generate complex free eccentricities ($z_{\rm
  free}, z'_{\rm free}$) weighted by the intrinsic distribution
inferred above, with $\sigma=0.007$ (Eq. \ref{eq:psigma}).  Each pair
of eccentricities produces correction factors relating the true and
nominal masses (see below Eq. \ref{eq:mvp}).  Keeping only those
generated eccentricity-pairs that reproduce the observed TTV phases
($\phi_{\rm ttv}, \phi'_{\rm ttv}$) within the 68\% error bars, the
median correction factor then gives the median true density. This
procedure introduces an uncertainty which we quantify by the $68\%$
bounds in the distribution of correction factors.  For the total
uncertainty in density measurement, we add these bounds quadratically
to the error-bars in TTV amplitudes.  Results are listed in Table
\ref{tab:masses2}.  The ratio of median to nominal density depends on
the observed TTV phases, but is typically of order $0.5$. Even for
planets with TTV phases consistent with zero, the correction factor
falls below unity. This is because the most likely eccentricity value
lies not at zero, but at $e \sim \sigma$.

\begin{figure*}
\centerline{\includegraphics[width=0.85\textwidth,trim=20 160 50 120,clip]{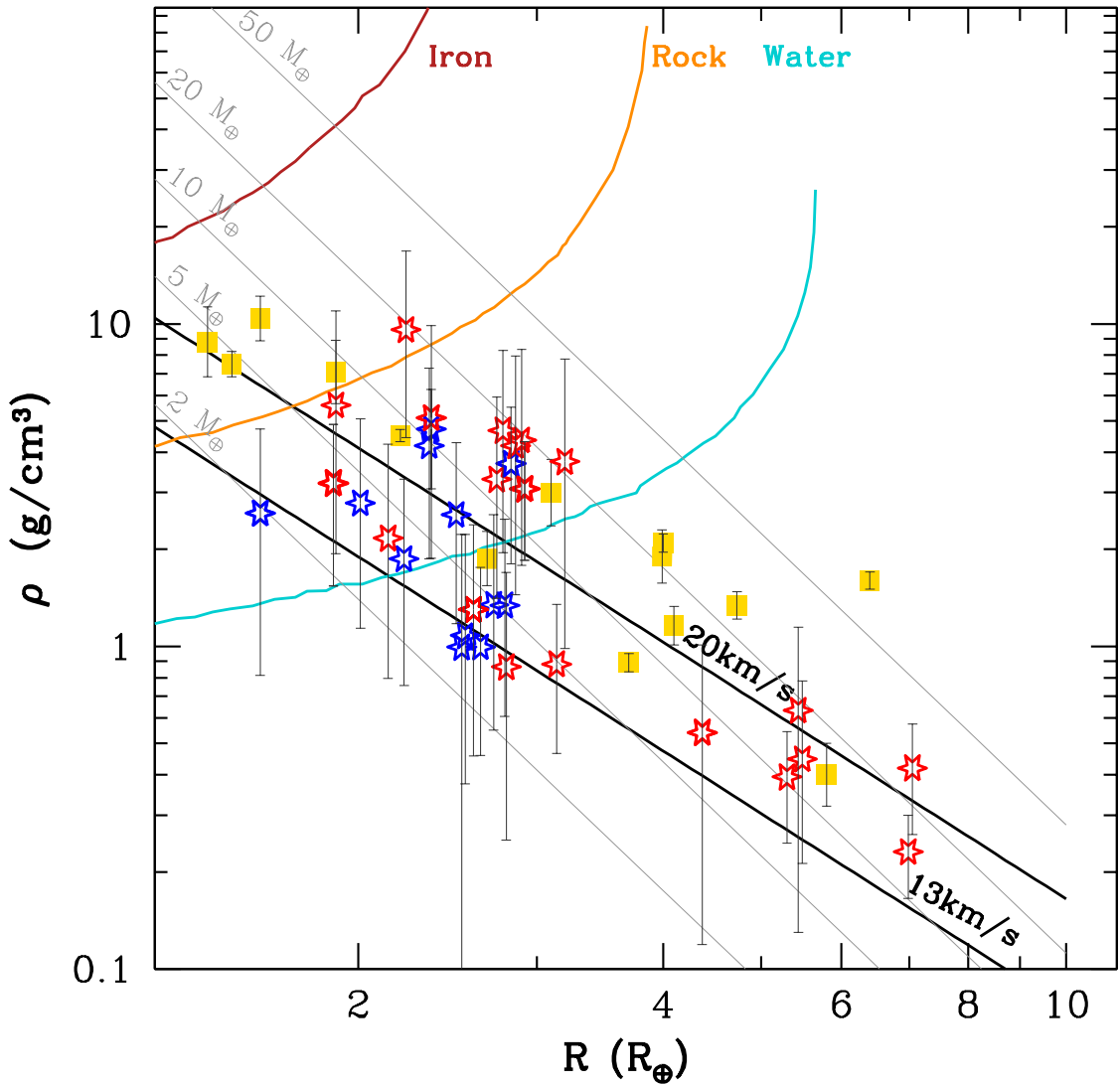}}
\caption{Similar to Fig. \ref{fig:ttv-density}, but showing the
  corrected densities for the $16$ low-e pairs, after adopting a
  Gaussian model for the eccentricity distribution with $\sigma=0.007$
  (Fig. \ref{fig:sigma}).  We apply no correction for Kepler 11b/c
  (koi 157.01/02).  Also shown here are density measurements from
  Table \ref{tab:masses} (golden squares, mostly by RV, except for
  Kepler 36b/c). The RV sample exhibits similar behavior to the TTV
  sample: larger planets are less dense, and are comparable in mass to
  the smaller ones. The points are best fit by the thick black line:
  $M = 3 M_\oplus (R/R_\oplus)$, corresponding to a constant surface
  escape velocity of $v_{\rm esc} = 20\km/\s$.  All planets satisfy
  $v_{\rm esc} \geq 13 \km/\s$, the sound speed of $10^4 \K$ hydrogen
  plasma.
All densities fall below  $\sim 15 \g/\cm^3$.
}
\label{fig:ttv-density2}
\end{figure*}

The median densities and their error-bars are plotted in
Fig. \ref{fig:ttv-density2}.  Combining our results with a sample of
low-mass planets that have previously determined masses (Table
\ref{tab:masses}, mostly through radial velocity), we find a best-fit
mass-radius relation of
\begin{equation}
  M \approx 3 M_\oplus\, \left( {R\over{R_\oplus}}\right)\, ,
\label{eq:fiducial}
\end{equation}
or a density $\rho \approx 3 \rho_\oplus (R/ R_\oplus)^{-2}$ where the
density of the Earth is $\rho_\oplus = 5.5\g/\cm^3$. This best-fit
differs from the one that applies for solar system planets
\citep{Lissaueretal11}, $M = (R/R_\oplus)^{2.06}\, M_\oplus$, or $\rho
\approx \rho_\oplus (R/R_\oplus)^{-1}$. 

{\y It is somewhat premature to discuss the significance of this
  difference, for a number of reasons.  First, we do not yet know what
  gives rise to the Solar System scaling, or if there is anything deep
  behind it. Second, our mass-radius relation is only an empirical fit
  to a select group of planets that are situated around $10$ days
  around their host stars, have radii $\geq 1 R_\oplus$ and masses
  $\sim 10-20 M_\odot$. It may not be universally applicable.  We do,
  however, note that the above mass-radius relation} 
corresponds to one that has a constant surface escape velocity of
$20\km/\s$. This is fortuitously close to the sound speed of hydrogen
plasma at $10^4\K$ ($13\km/\s$), suggesting that the process of
photoevaporation of hydrogen may be involved in some way.

In the next section, we discuss further details of our findings, as
well as their implications.

\section{Discussions}
\label{sec:correlation}

We first divide planets into two groups: `mid-sized' (those with $R
\geq 3 R_\oplus$) and `compact' (those smaller).  This division line
is partly inspired by an observation of the overall Kepler sample
\citepalias{Batalhaetal12}: there is a drastic fall-off in planet
numbers around $3 R_\oplus$ (right panel of
Fig. \ref{fig:kepler-starmass}), suggesting that a transition
occurs. {\y It is also partly inspired by the distinction in density
  between these two groups of planets: the 'mid-sized' ones are
  lighter than water planets, while 'compact' ones are of order or
  denser.}  We aim to discover the nature of this transition.

\subsection{Dynamical Excitation}
\label{subsec:eccdiscuss}

For the low-eccentricity population, we obtain an RMS eccentricity of
$e \approx 0.01$. Although small, this value is unexpected. To explain
the asymmetric pile-up of planet pairs on the far side of mean-motion
resonances, \citet{rr} and \cite{Batygin} argue that Kepler planets
have experienced protracted orbital damping. This could arise, for
instance, during their tenure living in disks. If this explanation is
correct, the small free eccentricities observed in TTV pairs would
require one to invoke a new source of eccentricity in the system to
re-excite the planets.

Eccentricities for the high-e population can be estimated assuming
that the true densities for these pairs follow the fiducial
mass-radius relation (Eq. \ref{eq:fiducial}). This yields $e\approx
|z_{\rm free}| \approx (\rho_{\rm nom}/\rho_{\rm fiducial})\, |\Delta|
\approx 0.1 - 0.2$.

Such high values are puzzling.  To understand their origin, we look
for correlations between eccentricity and other properties.  We do not
observe significant correlation with stellar mass, planet size or
planet period. In particular, mid-sized planets and compact planets
alike contain high-e pairs. We do note that all high-e pairs reside
near the 2:1 resonance.  Had they been near 3:2 or even closer
resonances, they might be unstable. We speculate that, while $25\%$ of
our pairs have high-e, the process that excites the eccentricities to
such high values could have happened more often, but only those near
2:1 remain.
%

We consider the hypothesis that these high-e population arise
  from planet-planet scattering. If so, one expects that the relative
inclinations between planets to be high, since close-encounters tend
to excite inclinations as well as eccentricities. However, Koi 152,
775, 829 each show 3 transiting planets.\footnote{Intriguingly, the
  third planets in these systems all lie close to 2:1 MMR with one of
  the TTV planets.}  Such a high incidence of transiting planets
argues against high relative inclination in the system, and hence
against planet scattering.

The presence of free eccentricity allows one to place a lower bound on
the tidal quality factor {\y in the planet}\footnote{We assume that
  dissipation in the star is irrelevant.} \citep[Q,
e.g.,][]{GoldreichSoter}.  Considering the high-e pair koi 1270.01/02,
for example, in order for its inner planet to retain eccentricity, it
needs to have a weak tidal dissipation with $Q \geq 1000$, approaching
the $Q$ values estimated for gas-rich planets in our Solar system ($Q
\sim 10^5$). In fact, if tidal dissipation in these planets is as
efficient as that in Earth (tidal quality factor $Q \approx 10$),
tidal interaction between planets and the host stars should have
reduced most planets' eccentricities to zero.

\subsection{Mid-sized planets, $R \geq 3 R_\oplus$}
\label{subsec:photo}

\subsubsection{Structure, Photoevaporation}
\label{subsubsec:detail1}

\begin{figure}
\centerline{\includegraphics[width=0.49\textwidth,trim=20 170 50 120,clip]{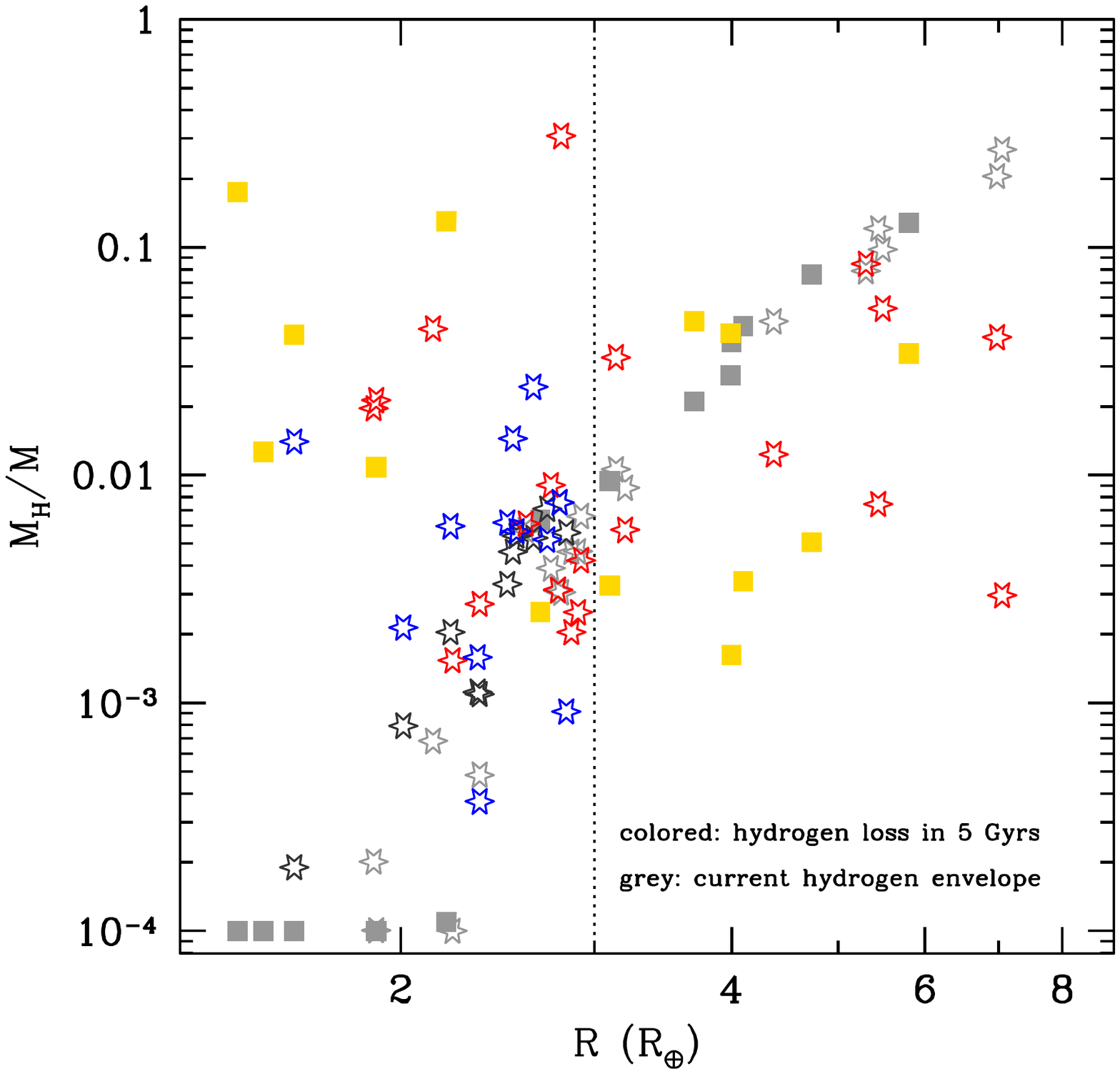}}
\caption{Hydrogen mass fractions on planets today (grey points) and
  those that are lost by 5 Gyr of photoevaporation (colored), for
  objects in Fig. \ref{fig:ttv-density2}. The grey points assume the
  planet is composed of a rocky core covered by a hydrogen atmosphere
  of solar composition, and the colored points assumes the hydrogen is
  eroded by energy-limited escape (Eq. \ref{eq:mdotenergy}). Given
    uncertainties in radius measurement, we are only sensitive to
    hydrogen mass fraction $\geq 10^{-3}$. All grey symbols falling
    below this value should be regarded as upper limits. For a given
  planet, if its grey symbol sits above its coloured one, it could
  have retained much of its hydrogen envelope over its lifetime.
  While if its grey symbol  sits lower, as is the case  for most
  compact objects, the planet should be a bare core.
%
}
\label{fig:evaporate-small}
\end{figure}

Mid-sized planets are a minority. They comprise $23\%$ of the total
Kepler planet candidates.\footnote{Here, we only include candidates in
  systems with multiple transiting objects, as they less polluted by
  false-positives \citep{keplerreal}.}  
Figure \ref{fig:ttv-density2} shows that mid-sized planets are less
dense than water and must have extensive H/He envelopes.  Based both
on the evolutionary calculations of \citet{Fortney}, and on our simple
model (see below), we conclude that the gas envelopes of these planets
are comparable in mass to their solid cores \citep[also see][for RV
planets]{gj436,hatp11,baraffe08}. {\y The actual composition of their
  solid cores does not affect this conclusion as the core sizes are
  small compared to the planet sizes.}

We calculate the fractional masses in their hydrogen envelopes that is
required to account for their observed densities and sizes. We assume
that the cores of these planets are made of pure rock, with the core
radii and masses related by Eq. (7) of \citet{Fortney}.  We assume the
photospheres of all planets are at a pressure of 1 bar, and a
temperature that is solely determined by irradiation, $T = T_{\rm
  eq}$, where $T_{\rm eq} = T_* (R_*/2a)^{1/2}$ is the equilibrium
temperature of a black-body at distance $a$ around a star with
temperature $T_*$ and radius $R_*$.  We integrate the thermal
structure of a solar-composition gas inward towards the core surface,
using the equation of state from \citet{SCVH}, and opacity from
\citet{Allard}. The temperature structure we obtain is nearly
isothermal near the surface and adiabatic in the deep interior. The
core mass is adjusted until the observed total mass is reproduced.
The result depends on the assumed internal luminosity since the latter
affects the temperature gradient and therefore the pressure scale
height at the base of the envelope.  For a luminosity of $10^{-11}
L_\odot$ (the current value on Neptune),
the mid-sized planets require hydrogen envelopes of order $2-30\%$ in
mass (Fig. \ref{fig:evaporate-small}).  The fractions rise to
$10-90\%$ when the internal luminosity drops by a factor of $100$.

The mid-sized planets exhibit surface escape velocities comparable to
the sound speed of an HII region ($10^4\K$, $13\km/s$), suggesting
that photoevaporation is important for their evolution. This is
corroborated by the absence of mid-sized planets with periods shorter
than a few days (see Fig. \ref{fig:kepler-pt}).  Here, we investigate
this possibility quantitatively.

We estimate mass-loss due to photoevaporation following \citet{Ruth}
for XUV photoevaporation. 
{\y For more detailed treatments see \citet{Owen,Lammer13}.}  Let the
stellar XUV luminosity be $L_{\rm UV}$. When this flux is absorbed by
the planetary atmosphere, it drives a thermal expansion and
outflow. In the limit of low flux,
the outflow rate can be approximated by the so-called energy-limited
formula,
\begin{eqnarray} {\dot M} & \approx &
\epsilon {{L_{\rm UV} \pi R_p^2}\over{4
      \pi a^2 \, v_{\rm esc}^2}}\,
  \sim  1.2\times 10^{10} \g/\s \left({{L_{\rm UV}}\over{5\times 10^{-6} L_\odot}}\right)\, \nonumber \\
  & & 
\times  \left({{0.1 {\rm AU}}\over{a}}\right)^2 \left({{R_p}\over{5
        R_\oplus}}\right)^3 \left({{15 M_{\oplus}}\over{M_p}}\right)\, ,
\label{eq:mdotenergy}
\end{eqnarray}
where we have scaled the XUV luminosity by $5\times 10^{-6} L_\odot$,
roughly the current solar value, and have assumed an efficiency
($\epsilon$) of unity.\footnote{We also equate the absorption cross
  section to the size of the planet disc because the atmosphere
  pressure scale height is very small, $H/R_p \approx 0.01
  \left({{R_p}/{2 R_\oplus}}\right)\, \left({{15
        M_{\oplus}}/{M_p}}\right)\, \left({T/{ 1000\K}}\right) \ll 1$.
} We have verified that the radiation/recombination-limited mass-loss
rates, relevant when heat loss is significant \citep{Ruth}, do not
apply: they lie above the energy-limited rate for all our objects
(compact and mid-sized).  We further note that the formally defined
sonic radius, $r_s = G M_p/2 c_s^2$ \citep{Parker}, frequently lies
inside the physical radius for the low surface-escape planets we
consider here.

Estimates for the fractional mass loss, integrated over 5 Gyrs, are
presented in Fig. \ref{fig:evaporate-small}. We have uniformly taken
the XUV luminosity to be $5\times 10^{-6}$ of the bolometric value,
and have adopted values for stellar radius and effective temperature
from the B12 
catalogue for the TTV planets {\y (see \S \ref{subsec:starsize})}, and
from discovery papers for planets in Table
\ref{tab:masses}).\footnote{Some of the stars have only recently
  become sub-giants but we ignore this complication here} We find that
mid-sized planets could shed between a few percent to a few tenths of
their masses, comparable to the hydrogen mass fractions inferred to be
present on these planets (grey symbols in
Fig. \ref{fig:evaporate-small}).  Our estimates for the mass-loss rate
are for the current systems. The rates were likely higher in the past
when the stars were more chromospherically active and brighter in the
FUV\citep[see, e.g.][]{Lammer,Ribas}, and also when the planets were
younger and fluffier \citep[also see][]{lopez}. 
{\y However, this active phase typically lasts $\sim 100$ Myrs, a
  short interval compared to the stellar age.}
{\y In addition,} the fractional XUV flux may be higher for stars
lower in mass than the Sun.  We therefore conclude that
photoevaporation has significantly sculpted these mid-sized
planets. {\y More detailed photoevaporation calculations are presented
  in \citet{owenwu}.}


In the following, we further argue that photoevaporation may explain
the mass-radius relation for mid-sized planets
(eq. \ref{eq:fiducial}), and naturally produces planets with
comparable core/envelope masses \citep[also see discussions in][]{lopez}.

Photoevaporation proceeds healthily as described by
Eq. \refnew{eq:mdotenergy} as long as the surface escape velocity of
the planet remains low. When the latter rises, the sonic radius (for a
wind of ionized hydrogen) moves much beyond the planet surface. This
throttles the photoevaporation rate as the
radiation/recombination-limited mass-loss, now relevant, decreases
exponentially with the expansion of the sonic surface \citep{Ruth}.
%
Now imagine a gas-rich planet that was exposed to the strong XUV flux
from its young host star before it has thermally contracted, either
because it was formed in-situ or was migrated very quickly 
\citep[eq. \refnew{eq:mdotenergy}, also see][]{outgas}. If such a
planet started with much more gas than that in its core, evolutionary
calculations \citep{Fortney} show that its size will expand during
mass loss, leading to further evaporation. This process continues
until mass of the solid core contributes significantly to gravity. At
this point, the planet shrinks when it loses mass until the escape
speed from its surface exceeds the thermal speed of ionized
hydrogen. Planets thus sculpted are expected to have envelopes that
are lower or comparable in mass to their cores. These arguments
highlight the need for a more accurate model, where one accounts for
the time evolution of both the XUV luminosity and planet structure {\y
  \citep[see, e.g.][]{Owen}.}

Regardless of the evolution, the mid-sized planets almost certainly
acquired their massive hydrogen envelopes by accreting from
their proto-planetary disks.

\subsubsection{Correlation with Stellar Mass}
\label{subsubsec:stars}

\begin{figure}
\centerline{\includegraphics[width=0.49\textwidth,trim=30 160 50 280,clip]{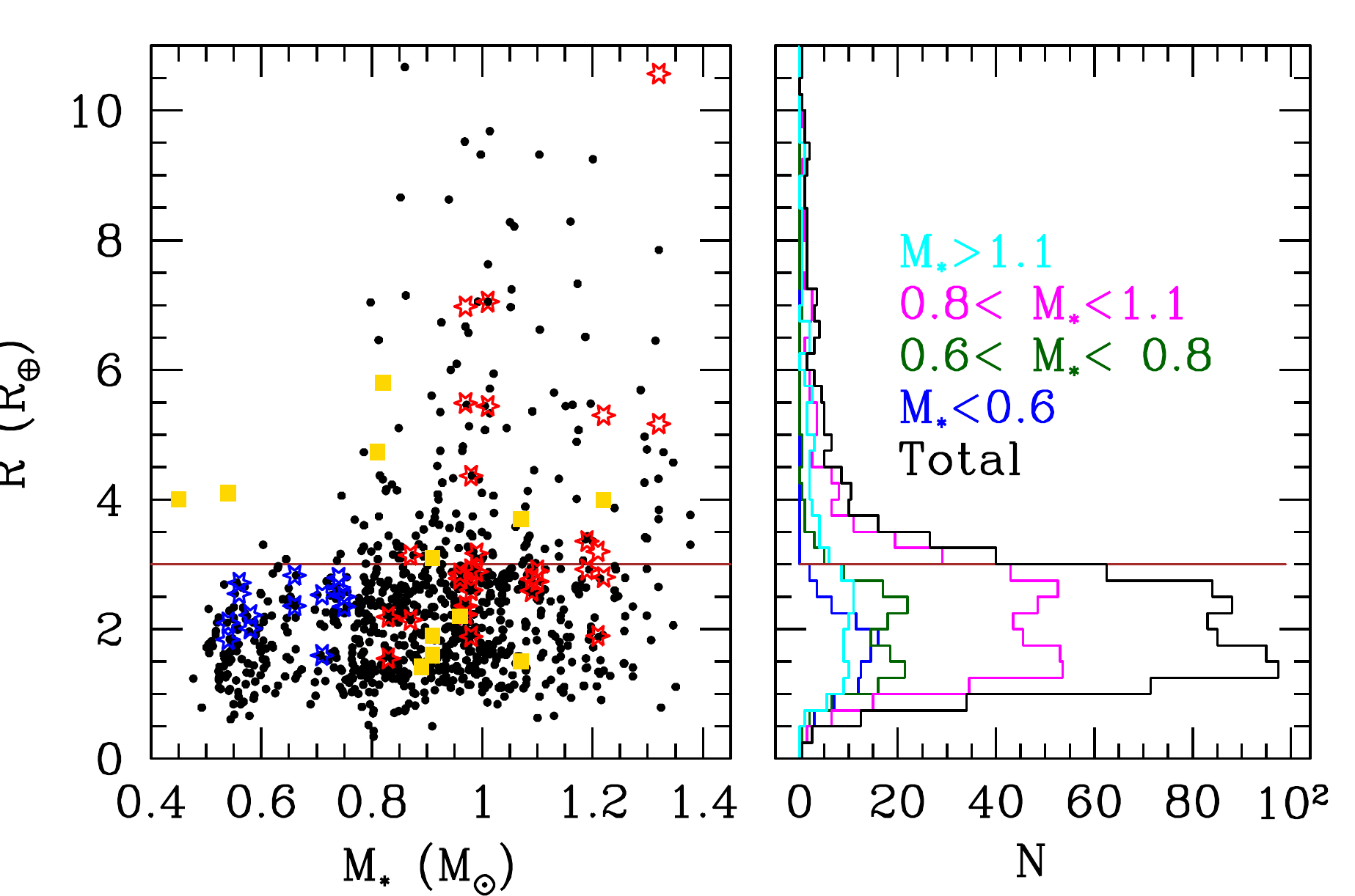}}
\caption{Size of Kepler planets versus their host mass (left panel)
  and distribution of planet sizes (right panel).  Here, we include
  only planet candidates from the B12 catalog that are in multiple
systems (black and colored points, where the latter are the sample
considered in this paper). The black histogram is the total size
distribution, and the colored curves are break-downs into different
stellar mass bins.  Overall, there is a precipitous drop of planet
numbers at the dividing line at $3 R_\oplus$. And mid-sized planets
are absent around stars less massive than $0.8 M_\odot$ (with the
exceptions of GJ 436b and GJ 3470b, both transiting planets discovered
by ground-based telescopes).  As the stellar mass increases from $0.4$
to $1.4 M_\odot$, the fraction of mid-sized planets rises from $0$ to
unity.
}
\label{fig:kepler-starmass}
\end{figure}

We focus here on one remarkable correlation between mid-sized planets
and their host properties, discovered in the published Kepler
catalogue.  In Fig. \ref{fig:kepler-starmass}, we display the
distribution of planet sizes as a function of their host
mass. Mid-sized planets occur almost exclusively around stars more
massive than $0.8 M_\odot$, or $T_{\rm eff} \geq 5000\K$ on the main
sequence. Quantitatively, the fraction of mid-sized planets, $23\%$ in
the overall sample, is nearly zero for stars less massive than $0.8
M_\odot$, and of order half for stars more massive than $1.1 M_\odot$.
{\y In} contrast, compact planets are more equitably distributed
around stars of all spectral types, albeit with a trend that average
planet sizes decrease around lower mass stars.

{\y This result differs from that in \citet{Howard} where they
  concluded that, albeit with weak statistics, the occurrence of
  planets with radii larger than $4 R_\oplus$ does not correlate with
  Teff. The difference may stem from the fact that while they included
  all candidates into their study, Fig. \ref{fig:kepler-starmass} only
  profiles systems with multiple transiting planets.  We suggest that
  their conclusion may be contaminated by false positives which are
  more populous among single transiting systems. Further studies are
  required to settle this difference.}

The fact that K/M-dwarfs do not host mid-sized, gas-rich
planets\footnote{\citet{Gaidos} argued that some K/M-dwarfs in the KIC
  are misclassified giants. If true, the sizes of their planets would
  increase to mid-sized. However, this would also mean only sub-giants
  have mid-sized planets, which would be puzzling.}
has not been previously noticed.  We offer two possible explanations
but no final resolution. First, photoevaporation around lower mass
stars could have been more severe than around F/G dwarfs as a result
of more potent and more prolonged chromospheric activity. Second,
planetary cores may take longer to form around lower mass stars, so
that they cannot accrete gas before the protoplanetary disks
disperses. Both scenarios can also explain why compact planets tend to
be smaller around lower-mass stars.


\subsection{Compact planets, $R < 3 R_\oplus$}
\label{subsec:Teq}

We turn now to the compact planets. The measured densities of compact
planets range from $\sim 1$ to $\sim 10 \g/\cm^3$
(Fig. \ref{fig:ttv-density2}). The largest compact planets ($R \sim 2-
3 R_\oplus$) have densities that hover around the water line ($\rho
\sim 2 \g/\cm^3$). At any given size, planets span almost a decade in
density.  While the mid-sized planets are most likely cores overlaid
with massive H/He envelopes, the structure of compact planets is less
certain.  They can either be water-rock mixtures, or rocky cores
enveloped by a small amount of hydrogen, {\y or a mixture of the two.}
It is important to resolve this degeneracy because a water-laden
planet would favour the scenario where the planet was constructed
outside the ice-line and then migrated inwards; a rocky core would be
consistent with {\it in-situ} formation.  Although density
measurements of individual planets cannot break this inherent
degeneracy \citep[see, e.g.,][]{degeneracy}, we attempt to break it
with our statistical sample.  
Fortuitously, the sample straddles the region where hydrogen
photoevaporation can significantly alter planet sizes.

\subsubsection{Density Correlates with Temperature}
\label{subsubsec:correlate}

\begin{figure}
\centerline{\includegraphics[width=0.49\textwidth,trim=10 140 30 120,clip]{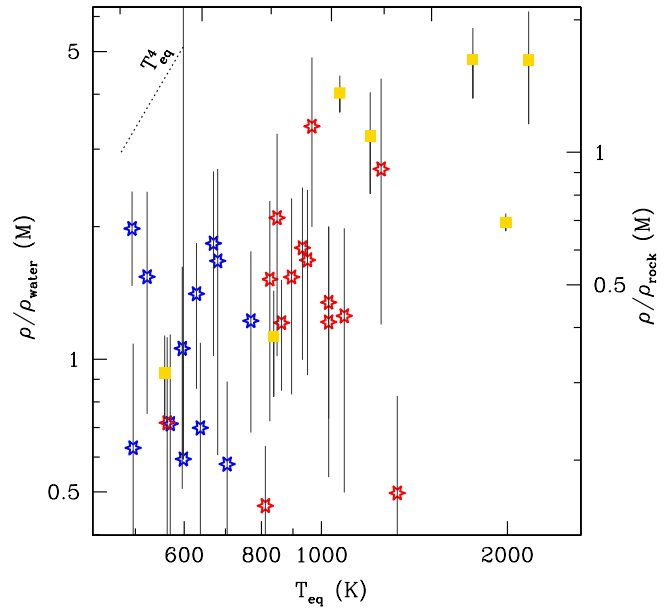}}
\caption{ Hotter compact planets are denser.  We scale the density of
  each compact planet to that of a water sphere of the same mass
  (left-hand ordinate), and to a rocky sphere (right-hand ordinate).
  Although the densities of watery and rocky spheres are
  mass-dependent, the conversion factor between the two is roughly
  constant (Fig. \ref{fig:ttv-density2}).  The horizontal axis is the
  black-body equilibrium temperature of the planet. We include all
  planets in Fig. \ref{fig:ttv-density2} that satisfy $R \leq 3.2
  R_\oplus$. While cool planets are roughly compatible with being
  water-worlds, the hottest ones have densities of rock.
  In contrast to density, planet masses show no systematic dependence
  on equilibrium temperature.}
\label{fig:mass-radius}
\end{figure}

Planet densities exhibit a remarkable correlation with environment:
planets with hotter equilibrium temperatures are in general denser
(Fig. \ref{fig:mass-radius}).  In that figure, we scale each planet's
density to the density of a pure-water (or pure-rock) planet of the
same mass (see the theory curves in Fig.
\ref{fig:ttv-density2}).\footnote{If we do not scale the densities,
  the correlation is still present but is more scattered.}  Cold
planets ($T_{\rm eq} \approx 600\K$) are compatible with being pure
water, while hot planets ($T_{\rm eq} \geq 1500 \K$) are compatible
with being pure rock.  Similar behavior has been noted by
\citet{lopez}.


\begin{figure}
\centerline{\includegraphics[width=0.48\textwidth,trim=20 160 50 120,clip]{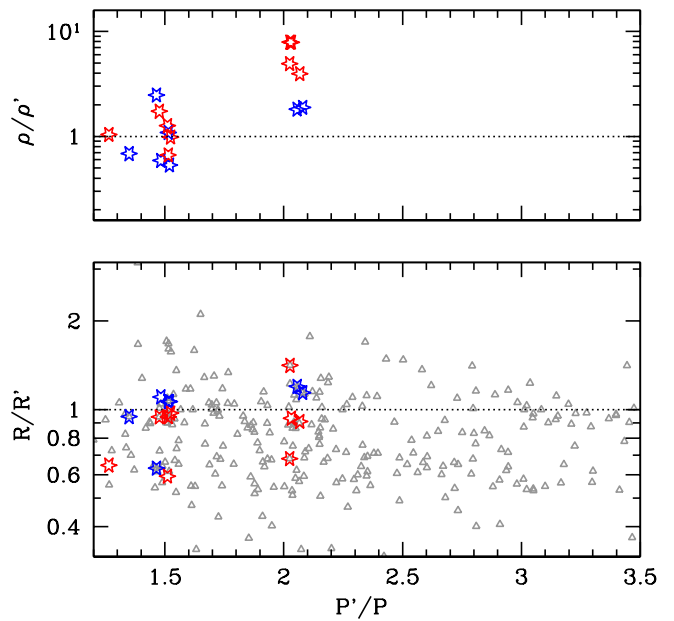}}
  \caption{Properties of planet pairs plotted versus the ratio of
    their orbital periods. The top panel shows the ratio of nominal
    density between the inner and outer planets for the compact
    planets in our sample.  For pairs that lie close to the 2:1 MMR,
    the inner planets are invariably denser, while pairs around more
    closely-spaced MMRs appear indistinguishable from each other. The
    bottom panel shows the size ratios for these pairs (colored
    points), as well as for all pairs in the \citetalias{Batalhaetal12}
    catalog that satisfy $P \leq 10$ days and where both components
    are compact (grey triangles). The inner planets tend to be
    smaller, more strikingly so when the planet pairs are spaced
    further apart.}
\label{fig:densitycontrast}
\end{figure}

A second piece of evidence for this correlation is furnished by
studying the density ratio within planet pairs (top panel of
Fig. \ref{fig:densitycontrast}). Unlike individual planet mass, the
mass ratio of two planets within a pair can be largely determined
without knowledge of the free eccentricity.
So here we simply plot the density contrast as the ratio of their
nominal densities. We find that the inner planets in pairs near 2:1
resonance are always denser by a factor of a few compared to the outer
ones, while pairs near closer resonances (5:4, 4:3, 3:2)  have
more comparable densities.\footnote{One notable
  exception is Kepler-36 where the inner planet is 8 times denser than
  the outer one, while the two are near 6:7 resonance
  \citep{kepler36}.}  
So some process appears to be at work to systematically compactify the
inner planets.

\begin{figure}
\centerline{\includegraphics[width=0.48\textwidth,trim=20 160 50 110,clip]{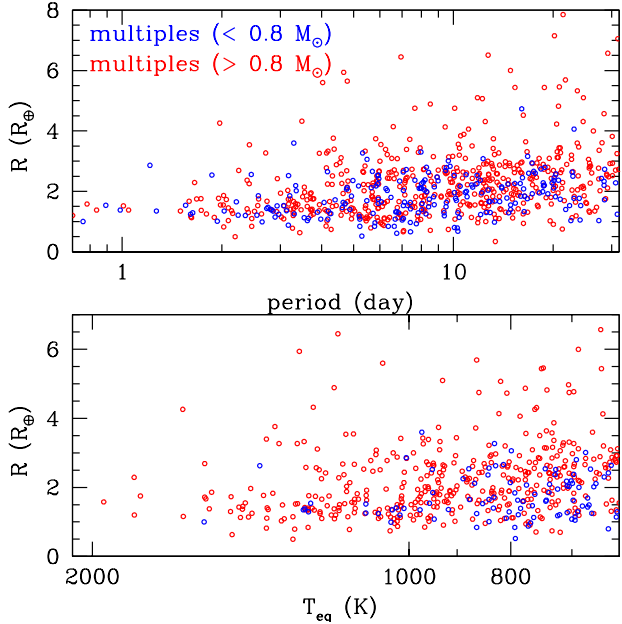}}
  \caption{Planet sizes in the \citetalias{Batalhaetal12} sample,
    plotted versus orbital period (above) and black-body equilibrium
    temperature (below). We only include systems with multiple
    candidates. Red points indicate those around more massive stars,
    and blue those around K \& M dwarfs.}
\label{fig:kepler-pt}
\end{figure}

The last piece of evidence comes from studying sizes of compact
planets in the \citetalias{Batalhaetal12} catalog. The bottom panel of
Fig. \ref{fig:densitycontrast} shows the size ratio within a planet
pair as a function of their period ratio. We only plot pairs with
inner period $P < 10$ days. Inner planets tend to be smaller than
their companions. This trend is more significant for more widely
spaced pairs. And it is absent for pairs orbiting too far from the
star ($P > 10$ days). We also plot planet sizes versus orbital period
or equilibrium temperature in Fig. \ref{fig:kepler-pt}. One observes
that compact planets further away from their stars can be larger than
those closer in \citepalias[also see, e.g., Fig. 8
of][]{Batalhaetal12}. The median size for planets inward of $5$ days
is $R \sim 1.5 R_\oplus$, compared with $R \sim 2.5 R_\oplus$ for
planets between $5$ and $20$ days.  These results suggest that planets
inward of $\sim 10$ days, or $\sim 1000\K$ around a sun-like star,
show signs of compactification.

{\y After the submission of this paper, \citet{Weiss} published a
  related study based on radial velocity data. They showed that, for
  planets in the mass range from a few to 20 $M_\oplus$, planet
  densities correlate with stellar incident flux. This is consistent
  with the correlation we discover here.}

\subsubsection{Rock or Water?}
\label{subsubsec:degeneracy}

A naive explanation for
Figs. \ref{fig:mass-radius}-\ref{fig:densitycontrast} is that all
compact planets started out as water/rock mixtures with water being
the dominant component, but the hottest ones have experienced total
water removal and became rocky spheres.

This, however, seems difficult on theoretical grounds.  It is
impossible to photoevaporate water (or any other high molecular weight
material).  Even if heated to $10^4 \K$, water cannot escape from the
surface of the observed planets, because its sound speed would be
$\sim 2$ km/sec, too low compared to the escape
velocity.\footnote{This sound speed may increase by a factor of a few
  if water is photodissociated and if both hydrogen and oxygen are
  ionized (E. Lopez, private communication). {\y However, this does
    not affect our conclusion qualitatively.}}
Since the mass loss rate is determined by the smaller of the
energy-limited and radiation/recombination-limited rates, and since
the latter is exponentially suppressed
when the sound speed is smaller than the escape speed \citep{Ruth},
the mass loss will be severely inhibited.\footnote{This argument means
  that the dusty evaporative clouds from KIC 12557548 \citep{eugene}
  is hard to explain unless the surface escape velocity from the
  planet is very low (E. Chiang, private communication).} This
conclusion differs from that of \citet{valencia} who adopt the
energy-limited mass-loss rate for water and hence find
significant water loss.

Another scenario, one that may have happened on Venus, is to
photodissociate water molecules into $H_2$ and oxygen and then
remove $H_2$ via either Jeans escape or hydrodynamical outflow driven
by photoevaporation.  This would require a reductive agent to absorb
the free oxygen.  If this reductive agent is atomic iron \citep[see,
e.g.][]{outgas0}, the original planet would have to contain $70\%$
iron and $30\%$ water in mass to absorb all free oxygen into $Fe_2
O_3$. Such a planet, at a mass of $10 M_\oplus$, would have a density
of $\sim 6 \g/\cm^3$ \citep[c.f. the theoretical curves in
Fig. \ref{fig:ttv-density},][]{Fortney}, much denser than the cold
planets we observe.
In addition, for the newly produced hydrogen to be photoevaporated, it
cannot be well mixed with the other heavy molecules. 
So we consider this option ineffective in compactifying Kepler
planets.

Instead we propose that the observed correlations can be explained if
{\bf all} compact planets started out as rocky cores overlaid with
{\y various} amounts of hydrogen. Planets experiencing the strongest
radiation would have their hydrogen completely depleted, exposing
dense rocky cores. We quantitatively investigate this possibility
here.

As a useful rule of thumb, at $T_{\rm eq} = 1000\K$, a hydrogen
envelope of $1\%$ in mass will expand the size of its planet by
$25-40\%$, depending on the core composition and
mass.\footnote{However, the expansion is much smaller if the hydrogen
  atmosphere is heavily polluted by metals.  Photoevaporation is also
  problematic in that case.} Another useful rule of thumb is the
timescale to erode $1\%$ of the planetary mass. We recast
eq. \refnew{eq:mdotenergy} as,
\begin{eqnarray}
{{1\% M_p}\over{\dot M}} & \approx &  {5 {\rm Gyrs}}\,
\left({{L_{\rm UV}}\over{5\times 10^{-6} L_\odot}}\right)^{-1}
\nonumber \\
& & 
\times \left({{0.1 {\rm AU}}\over{a}}\right)^{-2} \left({{R_p}\over{2.5
        R_\oplus}}\right)^{-3} \left({{7.5 M_\oplus}\over{M_p}}\right)^{-2}\, .
\label{eq:mdotenergy2}
\end{eqnarray}
where we have scaled the variables using typical parameters for
compact planets.  Since typical planets in our sample orbit with
period $\sim 10$ days, they just straddle the range where
photoevaporation can make an order-unity change to their radii.

For the compact planets, mass fractions in the hydrogen envelope run
from $0$ to $\sim 1\%$ (grey symbols in
Fig. \ref{fig:evaporate-small}, see assumptions for the calculations
in \S \ref{subsubsec:detail1}). Fig. \ref{fig:evaporate-small} also
shows that smaller compact planets in our sample could have
experienced evaporation loss that ranges from $1\%$ to $20\%$ in mass;
while larger compact planets typically suffer loss of order $1\%$ in
mass or lower. This then makes a self-consistent story. Larger planets
are larger because they can hold on to their hydrogen envelope, while
smaller ones have lost it all, mostly as a result of their close
proximity to the stars.

We can stretch the calculation a little further. Assuming the
planetary cores are not fully rocky (as assumed above) but are instead
made up of a half-water/half-rock mixture, we would then require
hydrogen mass fractions of order $10^{-3}$ or less to explain the
observed planets. Such a thin atmosphere have no chance of survival
for most of the planets in our sample. The situation is worse if the
cores are made up of pure water. We conclude that the cores of these
planets are likely composed of rock (or even denser material). This
statement is substantially weakened, however, if the internal
luminosity of the planet is much lower than that of Neptune.

So we are able to resolve the structural degeneracy in compact
planets, helped by the fact that typical planets in our sample have
$T_{\rm eq} \approx 1000\K$. Our proposal, that compact planets are
made of rocky cores overlaid with hydrogen $\leq 1\%$ in mass,
explains the rock-like density of hot planets (e.g., Kepler-10b,
CoRoT-7b, 55 Cnc e),
%
explains the factor of $\sim 2$ difference in size between hot and
cold compact planets in the Kepler catalogue, and explains the density
contrast within 2:1 pairs and the lack of density contrast in other
closer pairs.


The hydrogen mass fractions we infer for many of the compact planets
are of order $1\%$. This can be primordial in origin. Alternatively,
as suggested by \citet{outgas} and \citet{outgas0}, outgassing and the
subsequent break-down of water could produce a hydrogen envelope $\leq
1\%$ in mass. The former option implies a fast formation timescale,
while the latter implies a significant amount of primordial water
($\geq 10\%$ in mass) when the planet was formed.

\subsection{Rock or Water:  the \citet{lopez} study}

\citet{lopez} study the thermal evolution of Kepler-11b, employing the
energy-limited formula for photoevaporation.  Their stellar FUV flux
dims from $6\times 10^4$ erg/s/cm$^2$ at $100$ Myrs, to $37$
erg/s/cm$^2$ at $8$ Gyrs.  When they consider the case that Kepler-11b
is a rocky sphere overlaid with hydrogen, our preferred solution here,
they find that it initially had to have thirty times more hydrogen and
a much bigger size (for an efficiency $\epsilon=0.02$; see their
Fig. 2).  They also find that if the efficiency is 0.1, the initial
planet mass would have been so large as to destabilize the intricate
multi-planet system (Fig. 3 in that paper).  They therefore conclude
that Kepler-11b, with a density of $3\g/\cm^3$, is most likely a
water-world.

However, a concern with the \cite{lopez} study is that they always
adopt the energy-limited formula for mass-loss.
If the planet is initially $8 R_\oplus$ and if $\epsilon=0.02$, we
find that mass-loss transitions from energy-limited to
radiation/recombination-limited whenever the FUV flux exceeds $\approx
10^4$ erg/s/cm$^2$. This transition occurs at an FUV flux $40$
erg/s/cm$^2$ if $\epsilon=0.1$.  So the energy-limited formula
over-estimates the mass-loss at the critical early stage when the star
was FUV bright. Taking this into account may make the initial state of
Kepler-11b more probable under the no-water hypothesis than was made
out in \citet{lopez}.

To resolve these issues, it would be desirable to conduct an
evolutionary study of mass-loss that includes
radiation/recomination-limited mass loss rates. We leave such a study
to future work.

\subsection{Compact vs. Mid-sized}
\label{subsec:game}

We have suggested, based on the sharp drop-off in planet number at $R
= 3 R_\oplus$, that compact and mid-sized planets are physically
distinct categories. However the two types of planets span a similar
mass range and appear to form a continuum in their density and radius
(Fig. \ref{fig:ttv-density2}). It is unclear what induces one, but not
the other, to accrete a healthy gas envelope, when both types of
planets live in similar environments.
  We note that when a mid-sized and a compact planet coexist in a
  system, the former is always at least a factor of $2$ or more
  massive than the latter.
 
\section{Summary}
\label{sec:summary}

Based on two samples of low-mass planets--- $22$ planet pairs
characterized by the TTV method and a dozen planets characterized by
the radial velocity technique --- we reach the following conclusions:

\begin{enumerate}

\item TTV phases for most of our near-resonant pairs lie close to
  zero.  We conclude that the majority of these pairs are consistent
  with an eccentricity distribution that has a root-mean-squared value
  of $e \sim 0.01$. About a quarter of the pairs, on the other hand,
  have eccentricities as large as $0.1 - 0.4$.  True planet masses for
  the low-eccentricity population can be recovered statistically,
  based on their inferred eccentricity distribution.

\item Masses of planets are roughly proportional to their radii, such
  that the best-fit solution corresponds to a population that has a
  constant surface escape velocity of $20\km/\s$.

\item Mid-sized planets ($R \geq 3 R_\oplus$) in our sample invariably
  have such low densities that they have to contain substantial H/He
  envelopes. Their current rates of photoevaporation
  suggest that their masses and radii may have been limited by this
  process and that masses in their cores should be at least comparable
  to those of their gaseous envelopes.


\item Mid-sized planets, some $23\%$ of the \citetalias{Batalhaetal12}
  sample, show up exclusively around stars more massive than $0.8
  M_\odot$.
  Perhaps relatedly, planets around lower mass stars tend to have
  smaller sizes.  We do not have an explanation for these
  observations.

\item Densities of compact planets ($R \leq 3 R_\oplus$) fall between
  that of pure rock and pure water, with some even less dense than
  water. Planets with higher equilibrium temperatures tend to be
  denser and smaller.
  Planets in our sample fortunately straddle the region where
  photoevaporation could significantly erode a $1\%$ hydrogen
  envelope. This allows us to break the degeneracy in internal
  composition and show that it is possible to explain the observed
  correlations if these planets are mostly likely rocky cores overlaid
  with a layer of hydrogen that is $\leq 1\%$ in mass.  These planets
  are likely not water worlds.


\item The data suggest that $3 R_\oplus$ is a dividing line between
  `super-earths' (largely solid) and `hot Neptunes' (extensive gaseous
  envelopes). The atmospheres on mid-sized planets were most likely
  accreted, while those on compact planets may have come from
  accretion or outgassing.
  


%

\end{enumerate}

We foresee a number of directions to pursue in the future.  First,
more pairs are needed. We are currently restricted to pairs in the
immediate vicinity of resonances, and ones that orbit with periods
$5-12$ days.  Transit timing data of a longer span or a higher quality
would allow one to probe many more planet pairs. It will be of
interest to see whether the density-$T_{\rm eq}$ correlation
(Fig. \ref{fig:mass-radius}) extends to larger ranges.  More pairs
will also provide a more accurate eccentricity distribution.  At the
moment, uncertainties in mass determination arise both from
measurement errors in TTV amplitudes, and from statistical
uncertainties when converting nominal mass to genuine mass.

Our conclusion that gas-rich planets appear to have suffered
significant photoevaporation helps explain why there are so few of
them close to the star. It also predicts that more of these gas-rich
planets will show up at longer periods. The extended {\it Kepler} mission
should be able to resolve the issue. It is also of interest to follow
self-consistent thermal and photoevaporation evolution for these
planets, in order to determine their initial states. It remains
puzzling how such low-mass planets could have accreted so much
hydrogen, so close to the star. Lastly, the origin of hydrogen on
compact planets is an interesting question to pursue. If it was
accreted directly from the disk, these planets had to form before the
disks disperse. If it was outgassed from broken-down water, these
planets had to have water-rich interiors when they formed.

Most of our planets, gas-rich or compact, have masses $ \lesssim 20
M_{\oplus}$, a decade lower in mass than the Jovian planets at the
same period range.
This mass, incidentally, is roughly the gap-opening mass in a disk
with a scale height $H/R \approx 0.03$. This may be a valuable clue
for planet formation.

Moreover, other processes not considered in this study may be
important. We argue that an icy object cannot be converted into a
rocky ball. But we have not considered ice removal by bombardment of
planetesimals or proto-planets.

And lastly, we need to understand the source of eccentricity in
multiple low-mass planet systems.








\acknowledgments

We thank Doug Lin, Eugene Chiang, James Owen, Jonathan Fortney, Eric
Lopez and Daniel Huber for interesting discussions. YW acknowledges
the hospitality of KIAA-PKU, where part of this work was
performed. This research is supported by NSERC and the province of
Ontario, as well as NSF grant AST-1109776.  This study would not have
been possible without the amazing work by the Kepler team.


\bibliographystyle{apj}


\appendix

\begin{figure*}
\centering
\begin{tabular}{cc}
\includegraphics[width=0.4\textwidth,trim=20 130 20 80,clip]{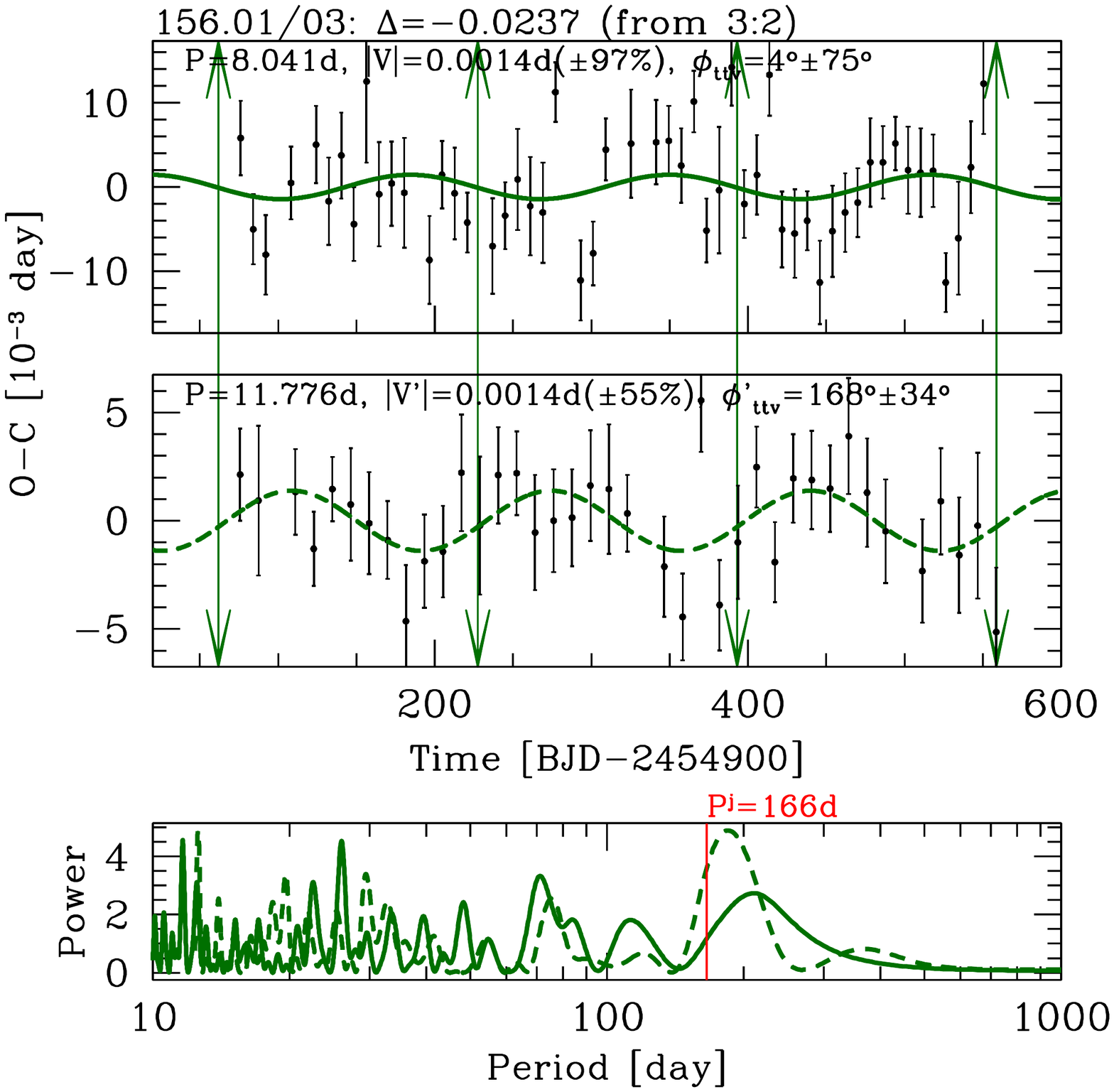} &
\includegraphics[width=0.4\textwidth,trim=20 130 20 80,clip]{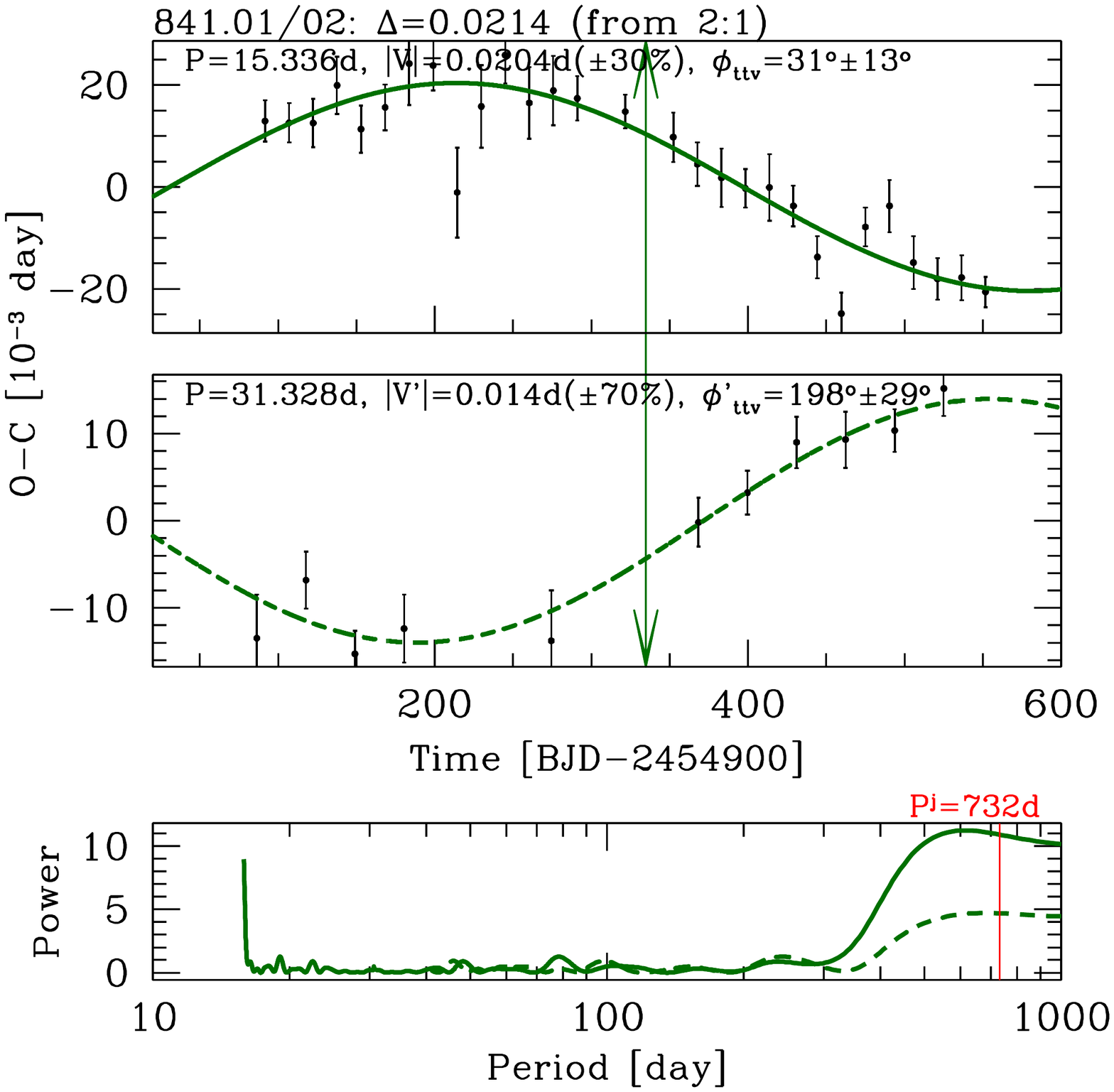}\\
\includegraphics[width=0.4\textwidth,trim=20 130 20 80,clip]{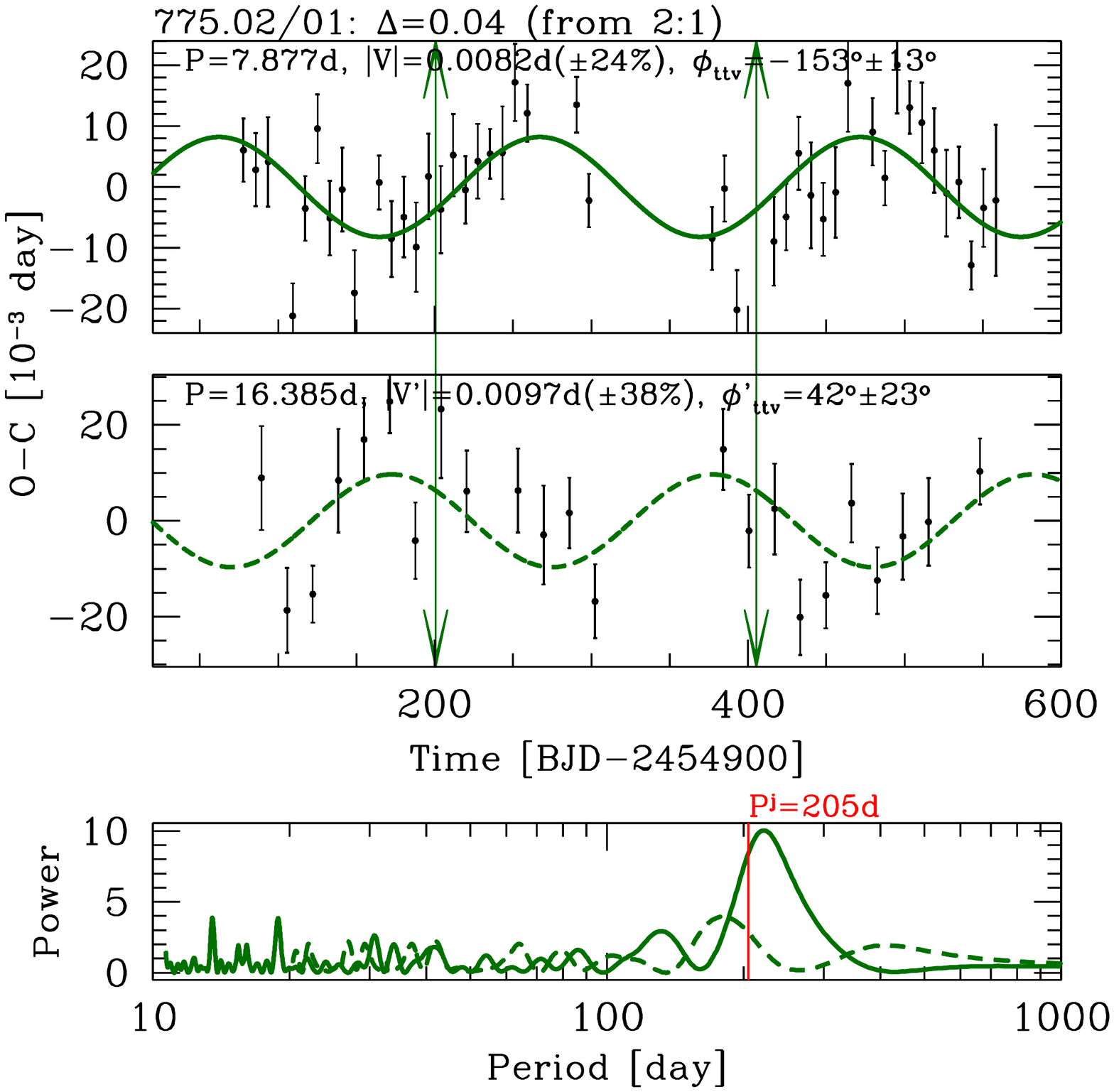} &
\includegraphics[width=0.4\textwidth,trim=20 130 20 80,clip]{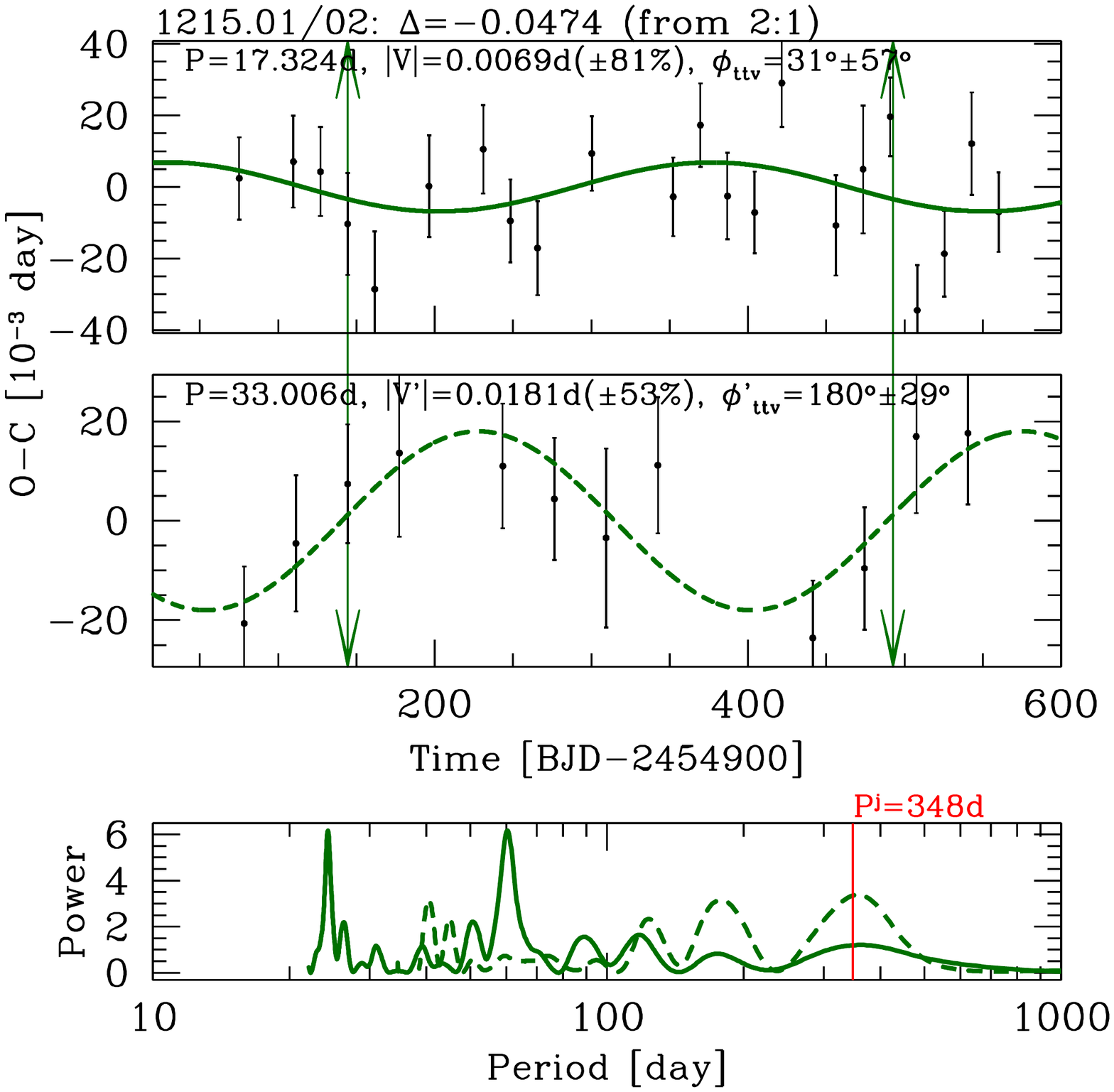}\\
\includegraphics[width=0.4\textwidth,trim=20 130 20 80,clip]{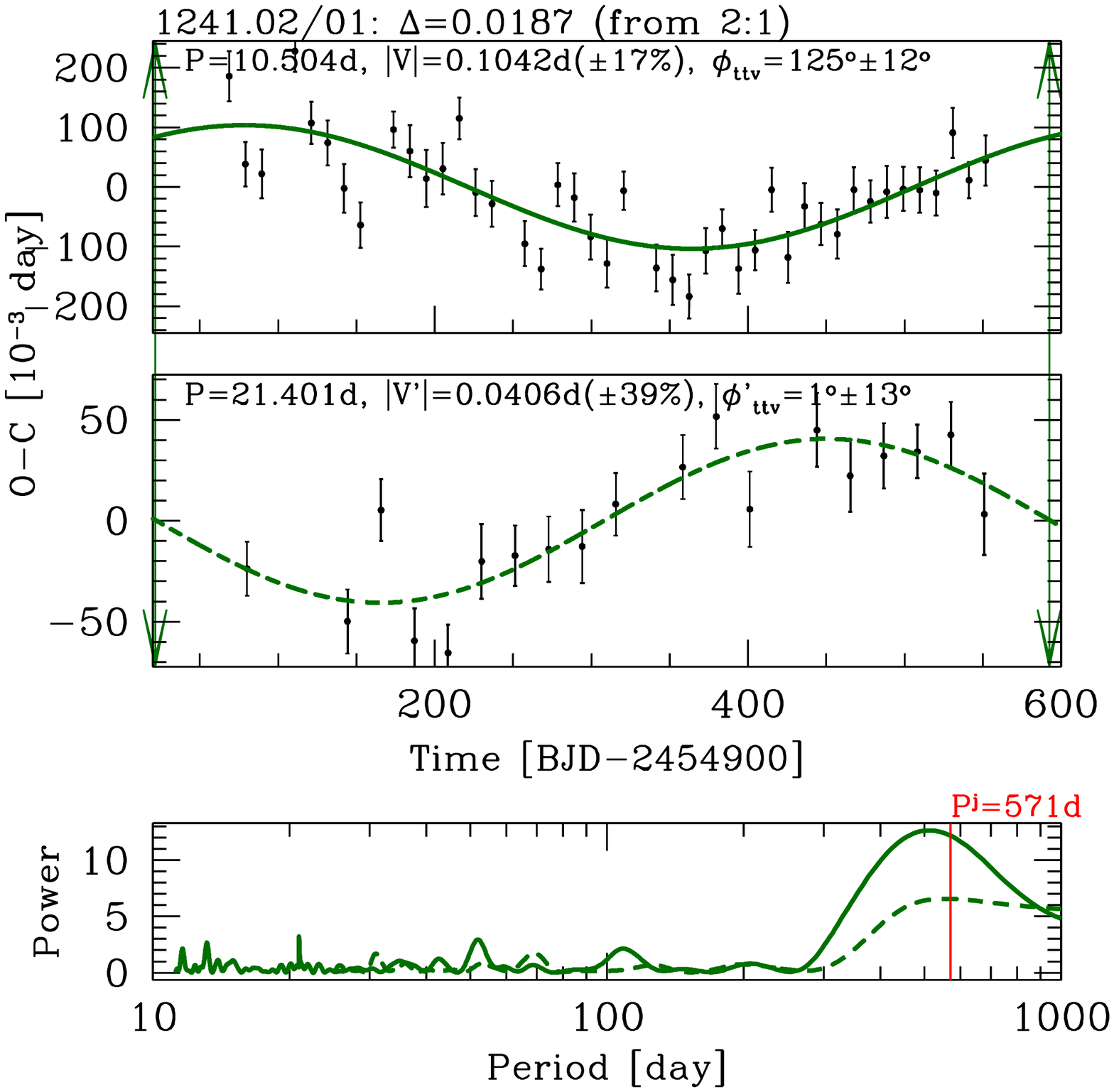} &
\includegraphics[width=0.4\textwidth,trim=20 130 20 80,clip]{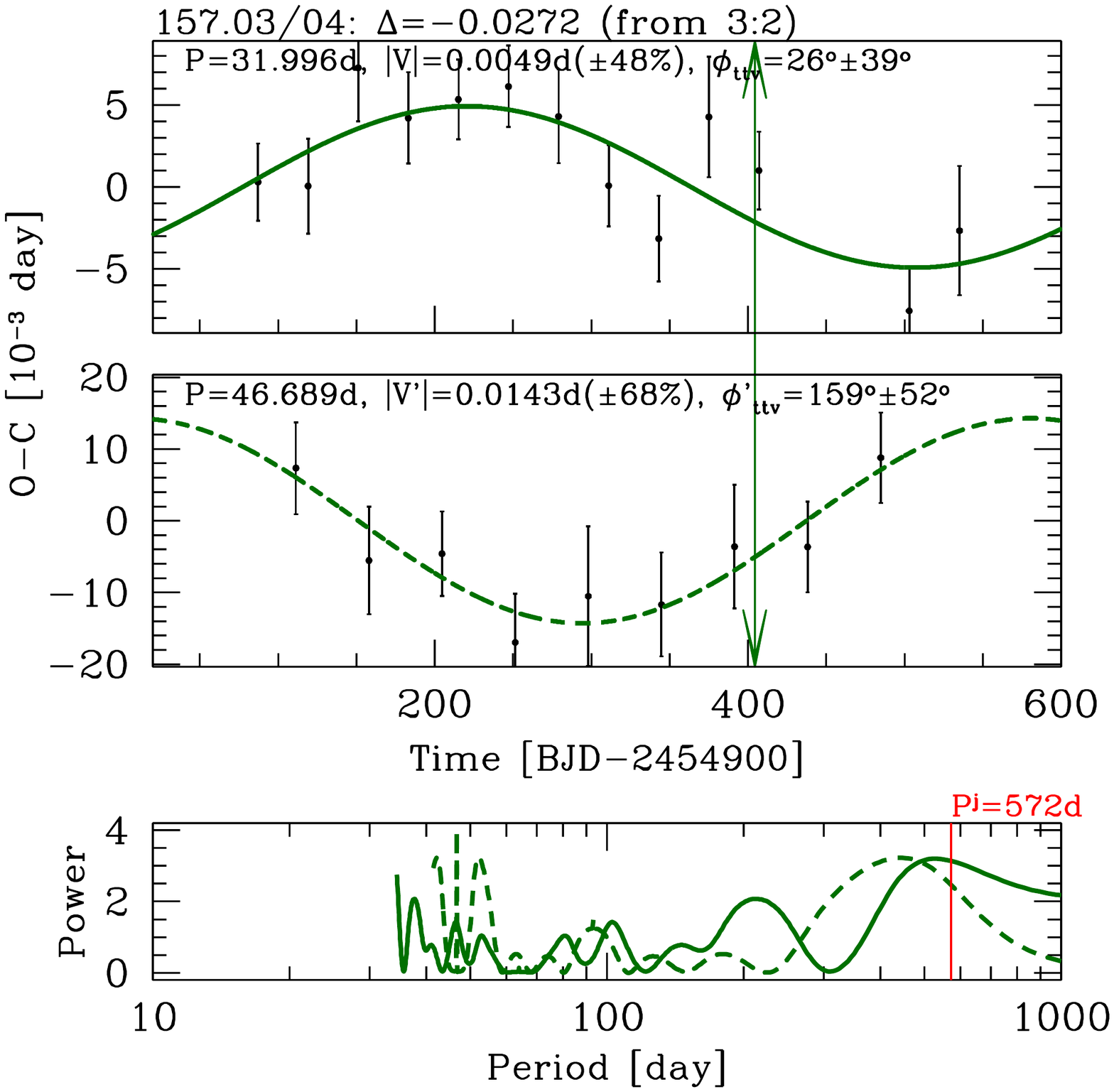} \\
\end{tabular}
\caption{TTV data and periodogram for 6 newly identified pairs of Kepler candidates. 
For each pair, the three panels show the inner and outer planet's TTV (top and middle panels),
and their periodograms (bottom).
 All
  planet candidates in this sample show clear sinusoidal variations at the
  desired super-period, marked by the red line in the 
  periodogram. 
In the top two panels, vertical arrows are times when the longitude
of conjunction points at the observer;
     green sinusoids are the best fit sinusoids at the super-period; and
     fitted amplitudes and phases (and 68\% confidence limits) are as listed, 
     where the phase is relative to the vertical arrows.
  The fitting procedure  is described in
  \citetalias{ttv1}. 
}
\label{fig:ttvs}
\end{figure*}

\end{document}